\documentclass[mathpazo]{cicp}

\usepackage{amsmath}
\usepackage{amsthm}
\usepackage{newtxmath}
\usepackage[T1]{fontenc}
\usepackage[utf8]{inputenc}
\usepackage{units}
\usepackage{graphicx}
\usepackage{microtype}
\usepackage{zymacros}

\DeclareMathOperator\supp{supp}
\makeatletter

\usepackage{changepage}

\usepackage{url}
\usepackage{booktabs}
\usepackage{amsfonts}
\usepackage{nicefrac}
\usepackage{caption}
\usepackage{xcolor}

\@ifundefined{showcaptionsetup}{}{%
 \PassOptionsToPackage{caption=false}{subfig}}
\usepackage{subfig}
\makeatother

\begin{document}

\title{Multi-scale Deep Neural Network (MscaleDNN) for Solving Poisson-Boltzmann Equation in Complex Domains}

\author[Liu Z Q et.~al.]{Ziqi Liu\affil{1}, Wei Cai\affil{2}, and Zhi-Qin John Xu\affil{3}\comma\corrauth}
\address{
\affilnum{1}\ Beijing Computational Science Research Center, Beijing, 100193, PR China \\
\affilnum{2}\ Dept. of Mathematics, Southern Methodist University, Dallas, TX 75275 \\
\affilnum{3}\ Institute of Natural Sciences, MOE-LSC and School of Mathematical Sciences, Shanghai Jiao Tong University, Shanghai, 200240, P.R. China }
\emails{{\tt liuziqi@csrc.ac.cn} (Z.~Q.~Liu), {\tt cai@smu.edu} (W.~Cai),          {\tt xuzhiqin@sjtu.edu.cn} (Z.~Q.~J.~Xu). Date: September 15, 2020, submitted to CiCP.}

\begin{abstract}
In this paper, we propose multi-scale deep neural networks (MscaleDNNs) using the idea of radial scaling in frequency domain and activation functions with compact support. The radial scaling converts the problem of approximation of high frequency contents of PDEs' solutions to a problem of learning about lower frequency functions, and the compact support activation functions facilitate the separation of frequency contents of the target function to be approximated by corresponding DNNs. As a result, the MscaleDNNs achieve fast uniform convergence over multiple scales. The proposed MscaleDNNs are shown to be superior to traditional fully connected DNNs and be an effective mesh-less numerical method for Poisson-Boltzmann equations with ample frequency contents over complex and singular domains.
\end{abstract}

\date{September 11, 2020}
\maketitle

\section{Introduction}
Deep neural network (DNN) has found many applications beyond its traditional applications such as image classification and speech recognition into the arena of scientific computing \cite{weinan2017deep, weinan2018deep, han2018solving,han2018deep, he2018relu, liao2019deep, raissi2019physics, hamilton2019dnn, strofer2019data,wang2020mesh}. However, to apply the commonly-used DNNs to computational science and engineering problems, we are faced with several challenges. The most prominent issue is that the DNN normally only handles data with low frequency content well, as shown by a Frequency Principle (F-Principle) that many DNNs learn the low frequency content of the data quickly with a good generalization error, but they are inadequate when high frequency data are involved \cite{xu_training_2018, rahaman2018spectral, xu2019frequency}. The fast convergence behavior of low frequency has been recently studied rigorously in theory in \cite{luo2019theory, zhang2019explicitizing, basri2019convergence, cao_towards_2020}. As a comparison, such a behavior of DNNs is the opposite of that of the popular multi-grid methods (MGM) for solving PDEs such as the Poisson-Boltzmann (PB) equation, where the convergence is achieved first in the high frequency spectrum of the solution due to the smoothing operations employed in the MGM. Considering the potential of DNNs in handling higher dimensional solutions and approximating functions without the need of a structured mesh as in traditional finite element or finite difference method, it is of great value to extend the capability of DNN as a mesh-less PDE solver. Therefore, it is imperative to improve the convergence of DNNs for solutions with fine structures as encountered in the electrostatic potentials of complex molecules.

The electrostatic interaction of bio-molecules with ionic solvents, governed by the Poisson-Boltzmann (PB) equation within the Debye-Huckel theory \cite{cai2013},  plays an important role in many applications including drug design and the study of disease. However, due to the complex surface structure of the bio-molecules, usually represented by a bead model, it has been a long outstanding challenging to design efficient numerical method to handle the singular molecular surface, which is either the van der Waals (vdW) surface being the sum of overlapping vdW spheres or the solvent accessible surface (SAS) generated by rolling a small ball on the vdW surface \cite{Lindskog},  and the complex distribution of the electrostatic potential over the molecular surfaces. Tradition finite element \cite{holst2001} and finite difference methods \cite{wei2007} have faced difficulties in the costly mesh generation and expensive solution of the discretized linear system. Therefore, in this paper, we will propose and investigate multi-scale DNNs, termed MscaleDNN, with the goal of approximating both low and high frequency information of a solution uniformly and developing a mesh-less solver for PDEs such as the PB equations in domains with complex and singular geometries.

Different learning behaviors among different frequencies are common. Leveraging this difference in designing neural network structure can benefit the learning process. In the field of computer vision, a series of works, such as image recovery \cite{deng2018learning}, super-resolution \cite{pan2018learning}, or classification \cite{wu2020multigrid}, have improved the learning performance, including the generalization and training speed, by utilizing the learning difference of different image frequencies. However, it should be noted that the frequency used in the computer vision tasks, is different from the response frequency of a mapping from the input (e.g., image) to the output (e.g., label), and the former refers to the frequency within an input (i.e. an image) with respect to the spatial locations inside the image. In this work, we address different response frequencies  of the mapping from the input to the output. As demonstrated in the previous work \cite{xu_training_2018}, the low response frequency is learned much faster than the high frequency.
The main idea of the MscaleDNN is to find a way to convert the learning or approximation of high frequency data to that of a low frequency one. Similar idea has been attempted in a previous work in the development of a phase shift DNN (PhaseDNN) \cite{cai2019phasednn}, where the high frequency component of the data was given a phase shift downward to a low frequency spectrum. The learning of the shifted data can be achieved with a small size DNN quickly, which was then shifted back (i.e., upward in frequency) to give approximation to the original high frequency data. The PhaseDNN has been shown to be very effective to handle highly oscillatory data from solutions of high frequency Helmholtz equations and functions of small dimensions. However, due to the number of phase shifts employed along each coordinate direction independently,  the PhaseDNN will result in many small DNNs and a considerable computational cost even for three dimensional problems. Here, we will consider a different approach to achieve the conversion of high frequency to lower one, namely, with a radial partition of the Fourier space, a scaling down operation will be used to convert higher frequency spectrum to a low frequency one before the learning is carried out with a small-sized DNN.  As the scaling operation only needs to be done along the radial direction in the Fourier space, this approach is easy to be implemented and gives an overall small number of DNNs, thus reducing the computational cost. In addition, borrowing the multi-resolution concept of wavelet approximation theory using compact scaling and wavelet functions \cite{daubechies1992ten}, we will modify the traditional global activation functions to ones with compact support. The compact support of the activation functions with sufficient smoothness will give a localization in the frequency domain where the scaling operation will effectively produce DNNs to approximate different frequency contents of a PDE solution. As a previous study shows \cite{shaham2018provable} that DNNs can approximate an intrinsically low dimensional function defined in a high dimensional space without the curse of dimensionality in terms of neuron number, provided it also has a sparse wavelet representation.
The proposed compact supported activation functions, similar to scaling and wavelet functions in the wavelet theory, will show their scale resolution capability in the MscaleDNNs.

Two types of MscaleDNN architectures are proposed, investigated, and compared for their performances. After various experiments, we demonstrate that MscaleDNNs solves elliptic PDEs much faster and can achieve a much smaller generalization error, compared with normal fully connected networks with similar overall size. We will apply MscaleDNNs to solve variable coefficient elliptic equations, including those solutions with a broad range of frequencies and over different types of domains such as a ring-shaped domain and a cubic domain with multiple holes. Also, to test the potential of MscaleDNN for finding Poisson-Boltzmann electrostatic solvation energy in bio-molecules, we apply MscaleDNN to solve elliptic equation with geometric singularities, such as cusps and self-intersecting surfaces in a molecular surface. These extensive experiments clearly demonstrate that the MscaleDNN is an efficient and easy-to-implement mesh-less PDE solver in complex domains.

The rest of the paper is organized as follows. In section 2, we will introduce frequency scaling to generate a MscaleDNN representation. Section 3 will present MscaleDNN structures with compact support activation functions. Section 4 will present a minimization approach through the Ritz energy  for finding the solutions of elliptic PDEs and a minimization approach through a least squared error for fitting functions. In section 5, we use two test problems to show the effectiveness of the proposed MscaleDNN over a normal fully connected DNN of same size. Next, numerical results of the solution of complex elliptic PDEs with complex domains by the proposed MscaleDNN will be given in Section 6. Finally, Section 7 gives a conclusion and some discussion for further work.

\section{Frequency scaled DNNs and compact activation functions}

In this section, we will first present a naive idea of how to use a frequency scaling in Fourier wave number space to reduce a high frequency learning problems for a function to a low frequency learning one for the DNN and will also point out the difficulties it may encounter as a practical algorithm.

Consider a band-limited function $f(\vx) \; \vx \in \mathbb{R}^{d}$, whose
Fourier transform $\widehat{f}(\mathbf{k})$ has a compact support, i.e.,%
\begin{equation}
\supp\widehat{f}(\mathbf{k})\subset B(K_{\text{max}})=\{\mathbf{k\in}%
\mathbb{R}^{d},|\mathbf{k|\leq}K_{\text{max}}\}.\label{support}%
\end{equation}

We will first partition the domain $B(K_{\text{max}})$ as union of $M$ concentric annulus with uniform or non-uniform width, e.g., for the case of uniform $K_0$-width
\begin{equation}
A_{i}=\{\mathbf{k\in}\mathbb{R}^{d},(i-1)K_{0}\leq|\mathbf{k|\leq}iK_{0}\},\quad K_{0}%
=K_{\text{max}}/M,\quad 1\leq i\leq M\label{anulnus}%
\end{equation}
so that %
\begin{equation}
B(K_{\text{max}})=%
{\displaystyle\bigcup\limits_{i=1}^{M}}
A_{i}.\label{part}%
\end{equation}

Now, we can decompose the function $\widehat{f}(\mathbf{k})$ as follows%
\begin{equation}
\widehat{f}(\mathbf{k})=%
{\displaystyle\sum\limits_{i=1}^{M}}
\chi_{A_{i}}(\mathbf{k})\widehat{f}(\mathbf{k})\triangleq%
{\displaystyle\sum\limits_{i=1}^{M}}
\widehat{f}_{i}(\mathbf{k}),\label{POU}%
\end{equation}
where $\chi_{A_{i}}$ is the indicator function of the set $A_{i}$ and
\begin{equation}
\supp\widehat{f}_{i}(\mathbf{k})\subset A_{i}.\label{SupAi}%
\end{equation}

\bigskip The decomposition in the Fourier space give a corresponding one in the physical space%
\begin{equation}
f(\vx)=%
{\displaystyle\sum\limits_{i=1}^{M}}
f_{i}(\vx),\label{Partx}%
\end{equation}
where %
\begin{equation}
f_{i}(\vx)=\mathcal{F}^{-1}[\widehat{f}_{i}(\mathbf{k})](\vx%
)=f(\vx)\ast\chi_{A_{i}}^{\vee}(\vx),\label{convol}%
\end{equation}
and the inverse Fourier transform of $\chi_{A_{i}}(\mathbf{k})$ is called the frequency selection kernel \cite{cai2019phasednn} and can be computed analytically using Bessel functions%
\begin{equation}
\chi_{A_{i}}^{\vee}(\vx)=\frac{1}{(2\pi)^{d/2}}
{\displaystyle\int\limits_{A_{i}}}
e^{i\mathbf{k\circ \vx}}dk.\label{kernel}%
\end{equation}

From (\ref{SupAi}), we can apply a simple down-scaling to convert the high frequency region $A_{i}$ to a low frequency region. Namely, we define a scaled version of $\widehat{f}_{i}(\mathbf{k})$ as
\begin{equation}
\widehat{f}_{i}^{(\text{scale})}(\mathbf{k})=\widehat{f}_{i}(\alpha
_{i}\mathbf{k}),\qquad \alpha_{i}>1,\label{fkscale}%
\end{equation}
 and, correspondingly in the physical space
\begin{equation}
f_{i}^{(\text{scale})}(\vx)=f_{i}(\frac{1}{\alpha_{i}}\vx%
),\label{fscale}%
\end{equation}
or
\begin{equation}
f_{i}(\vx)=f_{i}^{(\text{scale})}(\alpha_{i}\vx).
\label{fscale_inv}%
\end{equation}
We can see the low frequency spectrum of the scaled function $\widehat{f}_{i}^{(\text{scale})}(\mathbf{k})$ if $\alpha_i$ is chosen large enough, i.e.,
\begin{equation}
\supp\widehat{f}_{i}^{(\text{scale})}(\mathbf{k})\subset \{\mathbf{k\in}\mathbb{R}^{d},\frac{(i-1)K_{0}}{\alpha_{i}}\leq|\mathbf{k|\leq}\frac{iK_{0}}{\alpha_{i}}\}.\label{fssup}%
\end{equation}

Using the F-Principle of common DNNs \cite{xu2019frequency}, with $iK_{0}/\alpha_{i}$ being small, we can train a DNN $f_{\theta^{n_{i}}}(\vx)$, with $\theta^{n_{i}}$ denoting the DNN parameters, to learn $f_{i}^{(\text{scale})}(\vx)$ quickly
\begin{equation}
f_{i}^{(\text{scale})}(\vx)\sim f_{\theta^{n_{i}}}(\vx),\label{DNNi}
\end{equation}
which gives an approximation to $f_{i}(\vx)$ immediately%
\begin{equation}
f_{i}(\vx)\sim f_{\theta^{n_{i}}}(\alpha_{i}\vx)\label{fi_app}%
\end{equation}
and to $f(\vx)$ as well %
\begin{equation}
f(\vx)\sim%
{\displaystyle\sum\limits_{i=1}^{M}}
f_{\theta^{n_{i}}}(\alpha_{i}\vx).\label{f_app}%
\end{equation}

The difficulty of the above procedure used directly for approximating function and even more for finding a PDE solution is the need to compute the convolution in (\ref{convol}), which is computationally expensive for scattered data in the space, especially in higher dimensional problems. However, this framework will lay the structure for the multiscale DNN to be proposed next.

\section{MscaleDNN structures}
\subsection{Activation function with compact support}
In order to produce scale separation and identification capability of a MscaleDNN, we borrow the idea of compact scaling function in the wavelet theory \cite{daubechies1992ten}, and consider the activation functions with compact support as well. Compared with the normal activation function ${\rm ReLU}(x)=\max\{0,x\}$, we will see activation functions with compact support are more effective in MscaleDNNs. Two possible activation functions are defined as follows
\begin{equation}
\mathrm{sReLU}(x) ={\rm ReLU}(-(x-1))\times{\rm ReLU}(x)= (x)_{+} (1-x)_{+},
\end{equation}
and the quadratic B-spline with first continuous derivative
\begin{equation}
\phi(x) = (x-0)_{+}^2 - 3(x-1)_{+}^2 + 3(x-2)_{+}^2 - (x-3)_{+}^2,
\end{equation}
where $x_{+} = \max\{x, 0\}={\rm ReLU}(x)$, and the latter has an alternative form,
\begin{equation}
\phi(x) = {\rm ReLU}(x)^2 - 3{\rm ReLU}(x-1)^2 + 3{\rm ReLU}(x-2)^2 - {\rm ReLU}(x-3)^2.
\end{equation}

\begin{figure}[htbp]
\centering
\subfloat[ReLU]{\includegraphics[width=0.3\linewidth]{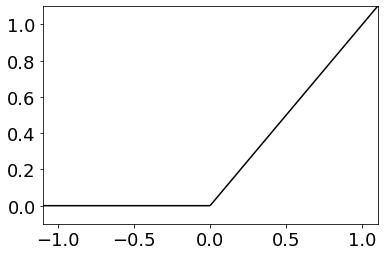}}
\subfloat[sReLU]{\includegraphics[width=0.3\linewidth]{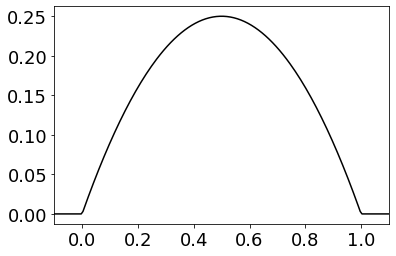}}
\subfloat[$\phi$]{\includegraphics[width=0.3\linewidth]{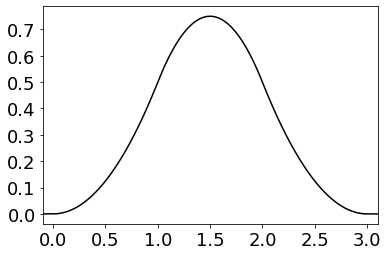}}
\caption{Activation functions in spatial domain.}
\label{actfunc}
\end{figure}

\begin{figure}[htbp]
\centering
\subfloat[sReLU]{\includegraphics[width=0.4\linewidth]{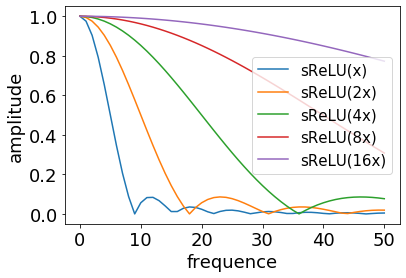}}
\subfloat[$\phi$]{\includegraphics[width=0.4\linewidth]{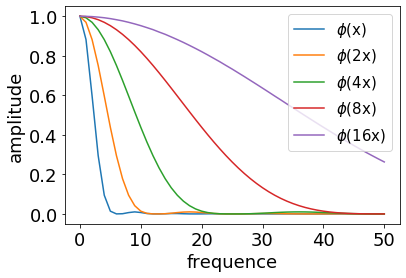}}
\caption{Activation functions in frequency domain, normalized by the maximum of each case.}
\label{freq}
\end{figure}

All three activation functions are illustrated in spatial domain in Fig. \ref{actfunc} and the Fourier transforms of both $\sReLU(x)$ and $\phi(x)$ are illustrated in Fig. \ref{freq}.

\begin{figure}[htbp]
\centering
\subfloat[MscaleDNN-1]{\includegraphics[width=0.45\linewidth]{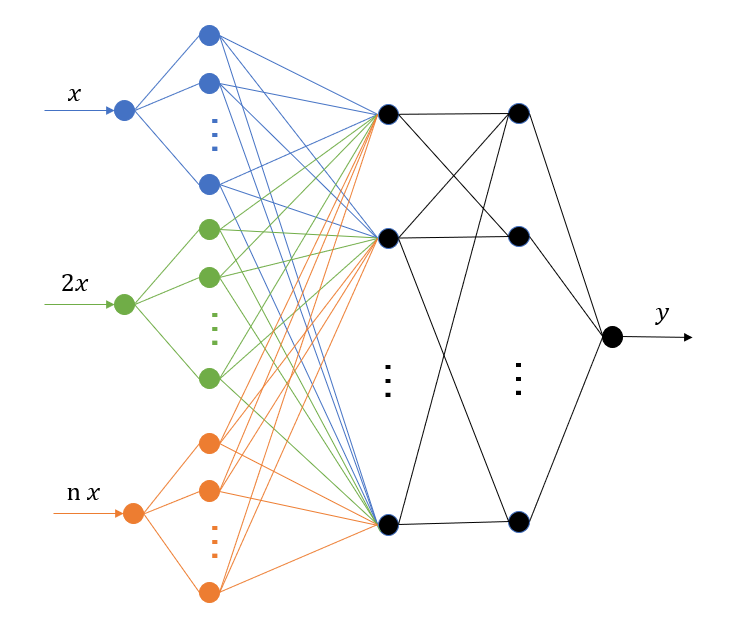}}
\subfloat[MscaleDNN-2]{\includegraphics[width=0.45\linewidth]{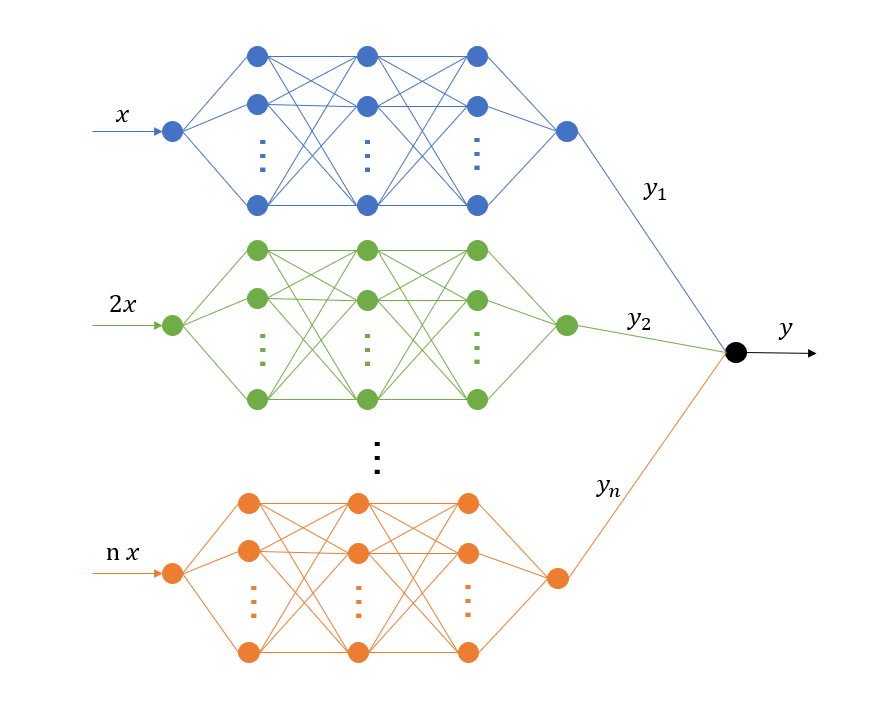}}
\caption{Illustration of two MscaleDNN structures.}
\label{net}
\end{figure}

\subsection{Two MscaleDNN structures}
While the procedure leading to (\ref{f_app}) is not practical for numerical approximation in high dimension, it does suggest a plausible form of function space for finding the solution more quickly with DNN functions. We can use a series of $a_{i}$ ranging from $1$ to a large number to produce a MscaleDNN structure to achieve our goal in speeding up the convergence for solution with a wide range of frequencies with uniform accuracy in frequencies. For this purpose, we propose the following two multi-scale structures.
\medskip

\noindent {\bf MscaleDNN-1} For the first kind, we separate the neuron in the first hidden-layer into to $N$ parts. The  neuron in the $i$-th part receives input $a_i\vx$, that is,  its output is $\sigma(a_i\vw\cdot\vx+b)$, where $\vw$, $\vx$, b are weight, input, and bias parameters, respectively. A complete MscaleDNNs takes the following form
\begin{equation}
    f_{\vtheta}(\vx) = \vW^{[L-1]} \sigma\circ(\cdots (\mW^{[1]} \sigma\circ(\mW^{[0]} (\vK\odot\vx) + \vb^{[0]} ) + \vb^{[1]} )\cdots)+\vb^{[L-1]}, \label{mscalednn}
\end{equation}
where $\vx\in\mathbb{R}^d$, $\mW^{[l]}\in\mathbb{R}^{m_{l+1}\times m_{l}}$, $m_l$ is the neuron number of $l$-th hidden layer, $m_0=d$, $\vb^{[l]}\in\mathbb{R}^{m_{l+1}}$,
$\sigma$ is a scalar function and ``$\circ$'' means entry-wise operation, $\odot$ is the Hadamard product and
\begin{equation}
\vK=(\underbrace{a_1,a_1,\cdots,a_1}_{\text{1st part}},\underbrace{a_2,a_2,\cdots,a_2}_{\text{2nd part}},a_3,\cdots,a_{i-1},\underbrace{a_i,a_i,\cdots,a_i}_{\text{ith part}},\cdots,\underbrace{a_{N},a_{N}\cdots,a_{N}}_{\text{Nth part}})^T,
\end{equation}
where $a_i=i$ or $a_i=2^{i-1}$.

We refer to this structure as Multi-scale DNN-1 (MscaleDNN-1) of the form in Eq. (\ref{mscalednn}), as depicted in Fig. \ref{net}(a).
\bigskip

\noindent {\bf MscaleDNN-2} A second kind of multi-scale DNN is given in Fig. \ref{net}(b), as a sum of $N$ subnetworks, in which each scale input goes through a subnetwork.  In MscaleDNN-2, weight matrices from $W^{[1]}$ to $W^{[L-1]}$ are block diagonal. Again, we could select the scale coefficient $a_i=i$ or $a_i=2^{i-1}$.

For comparison studies, we will define a ``{\bf normal}'' network as an one fully connected DNN with no multi-scale features. We would perform extensive numerical experiments to examine the effectiveness of different settings and use an efficient one to solve complex problems. All models are trained by Adam \cite{kingma2014adam} with learning rate $0.001$.

\section{MscaleDNN for approximations and elliptic PDE's solutions}
In this section, we will address two problems, i.e., fitting functions and solving PDEs such as the PB equations, to show the effectiveness of MscaleDNNs in the following sections.

\subsection{Mean squared error training for fitting functions}

A DNN, denoted by $f_{\theta}(\vx)$, will be trained with the mean squared error (MSE) loss to fit a target function $f^{*}(\vx)$. The loss function is defined as
\begin{equation}
\min L(f_{\theta}) = \int_\Omega |f^{*}(\vx) - f_{\theta}(\vx)|^2 \D \vx,
\end{equation}
where $f_{\theta}(\vx)$ is a neural network with parameter $\theta$.

In our training process, the training data are sampled from $f(\vx)$ at each training epoch, the loss at each epoch is
\begin{equation}
L_{S}(f_{\theta}) = \frac{1}{n} \sum_{x\in S} |f^{*}(\vx) - f_{\theta}(\vx)|^2,
\end{equation}
where $n$ is the sample size in $S$.

The above training process requires all information of the target function, which indicates such a training process is not of much practical use.  We conduct this study to examine the ability of a DNN in fitting high-frequency functions given sufficient information of the target function.

\subsection{A Ritz variational method for Poisson-Boltzmann equations}

Let us consider the following elliptic Poisson-Boltzmann equation,
\begin{equation}
-\nabla(\epsilon(\vx)\nabla^{}u(\vx))+\kappa(\vx)u(\vx)=f(\vx), \quad \vx \in\Omega\subset
\mathbb{R}^{d}, \label{prob}
\end{equation}
where $\epsilon(\vx)$ is the dielectric constant and $\kappa(\vx)$ the inverse Debye-Huckel length of an ionic solvent. For a typical solvation problem of a solute such as a bio-molecule in ionic solvent, the dielectric constant will be a discontinuous function across the solute-solvent interface $\Gamma$ where the following transmission condition will be imposed,
\begin{equation}
[u](\vx)  = 0, \quad \vx \in \Gamma ,
\label{continuouscond}
\end{equation}
\begin{equation}
[\epsilon \frac{\partial u}{\partial n}](\vx)   = 0,  \quad \vx \in \Gamma, \label{derivativecond}
\end{equation}
where $[ ]$ denotes the jump of the quantity inside the square bracket and, for simplicity, an approximate homogeneous boundary condition on $\partial \Omega$ is used for this study, i.e.
\begin{equation}
    u|_{\partial \Omega}=0.
\end{equation}

We will apply the deep Ritz method as proposed in \cite{weinan2018deep},  which produces a variational solution $u(\vx)$ of equation (\ref{prob}) and (\ref{continuouscond}) (\ref{derivativecond}) through the following minimization problem%
\begin{equation}
u=\arg\min_{v\in H^1_0(\Omega) }J(v),\label{Ritz}%
\end{equation}
where the energy functional is defined as%
\begin{equation}
J(v) =\int_{\Omega}
\frac{1}{2}\left(  \epsilon(\vx)|\nabla v(\vx)|^{2}+\kappa(\vx)v(\vx)^{2}\right)  d\vx -
{\displaystyle\int\limits_{\Omega}}
f(\vx)v(\vx)d\vx
 \triangleq%
{\displaystyle\int\limits_{\Omega}}
E(v(\vx)) \D\vx.\label{energy}
\end{equation}

We use the MscaleDNN $u_{\theta}(\vx)$ to represent trial functions $v(\vx)$ in the above variational problem, where $\theta$ is the DNN parameter set. Then, the MscaleDNN solution is
\begin{equation}
\vtheta_{\ast}=\arg\min_{\theta}J(u_{\theta}(\vx)).\label{Ritz-t}%
\end{equation}

The minimizer $\vtheta_{\ast}$ can be found by a stochastic gradient decent (SGD) method,%
\begin{equation}
\vtheta^{(n+1)}=\vtheta^{(n)}+\eta\nabla_{\vtheta%
}J(u_{\theta}(\vx)). \label{sgd}%
\end{equation}
The integral in Eq. (\ref{energy})  will only be sampled at some random points $\vx_i, i=1, \cdots, n$ at each training step  (see (2.11) in \cite{weinan2018deep}), namely,
\begin{align}
\nabla_{\vtheta}J(u_{\theta}(\vx))\sim\nabla_{\vtheta}\frac{1}{n}%
{\displaystyle\sum\limits_{i=1}^{n}}
E(u_{\theta}(\vx_i)).\label{GradJ}
\end{align}

At convergence $\vtheta^{(n)}\rightarrow\vtheta_{\ast}$, we obtain a MscaleDNN solution $u_{\theta_{\ast}}(\vx)$.

\medskip
\noindent{ \bf Variational functional for non-homogeneous Dirichlet boundary conditions}
\medskip

To derive the functional for (\ref{prob}) with a non-homogeneous boundary condition%
\begin{equation}
u|_{\partial\Omega}=g,
\end{equation}
we will construct a spatial extension function $\widetilde{g}\in
C^{2}(\Omega)$, such that%
\begin{equation}
\widetilde{g}|_{\partial\Omega}=g,\frac{\partial\widetilde{g}}{\partial
n}|_{\partial\Omega}=0,\ \sup(\widetilde{g})\cap\Gamma=\emptyset, \label{gExt}%
\end{equation}
and consider the function %
\begin{equation}
w=u-\widetilde{g},
\end{equation}
which satisfies equation (\ref{prob}) with a new right hand side
\begin{equation}
\widetilde{f}=f+\epsilon(x)\Delta\widetilde{g}-\kappa(x)\widetilde
{g}\label{ftilda}%
\end{equation}
with an homogeneous boundary condition on $\partial\Omega$, and can
be found as the minimizer of the following minimization problem%
\begin{equation}
w=\arg\min_{v\in H_{0}^{1}(\Omega)}J(v),
\end{equation}
where%
\begin{equation}
J(v)=J(v;\widetilde{f})=\frac{1}{2}%
{\displaystyle\int\limits_{\Omega}}
\left(  |\epsilon(x)\nabla v(x)|^{2}+\kappa(x)v(x)^{2}\right)  dx-%
{\displaystyle\int\limits_{\Omega}}
\widetilde{f}v(x)dx.
\end{equation}
Now consider the set
\begin{equation}
V_{g}=\{\varpi \in H^1(\Omega)  | \varpi=\widetilde{g}+v,v\in H_{0}^{1}(\Omega)\}=\widetilde{g}\oplus
H_{0}^{1}(\Omega) \subset H^1(\Omega).\label{Vg}%
\end{equation}

Using the definition of $V_{g}$ in (\ref{Vg})\ and $\widetilde{f}$ in
(\ref{ftilda}),  we can show that for piecewise constant $\epsilon(x),$ %
\begin{align*}
J(v)  & =\frac{1}{2}%
{\displaystyle\int\limits_{\Omega}}
\left(  |\epsilon(x)\left(  \nabla\varpi-\nabla\widetilde{g}\right)
|^{2}+\kappa(x)\left(  \varpi-\widetilde{g}\right)  ^{2}\right)  dx-%
{\displaystyle\int\limits_{\Omega}}
\left(  f+\epsilon(x)\Delta\widetilde{g}-\kappa(x)\widetilde{g}\right)
\left(  \varpi-\widetilde{g}\right)  dx\\
& =\left(  \frac{1}{2}%
{\displaystyle\int\limits_{\Omega}}
\left(  |\epsilon(x)\nabla\varpi|^{2}+\kappa(x)\varpi^{2}\right)  dx-%
{\displaystyle\int\limits_{\Omega}}
f\varpi dx\right)  -%
{\displaystyle\int\limits_{\Omega}}
\left(  \epsilon(x)\nabla\varpi\nabla\widetilde{g}+\kappa(x)\varpi
\widetilde{g}\right)  dx\\
& +\frac{1}{2}%
{\displaystyle\int\limits_{\Omega}}
\left(  |\epsilon(x)\nabla\widetilde{g}|^{2}+\kappa(x)\widetilde{g}%
^{2}\right)  dx-%
{\displaystyle\int\limits_{\Omega}}
\left(  \epsilon(x)\Delta\widetilde{g}-\kappa(x)\widetilde{g}\right)  \varpi
dx+%
{\displaystyle\int\limits_{\Omega}}
f\widetilde{g}dx+%
{\displaystyle\int\limits_{\Omega}}
\left(  \epsilon(x)\Delta\widetilde{g}-\kappa(x)\widetilde{g}\right)
\widetilde{g}dx\\
& =J(\varpi;f)-%
{\displaystyle\int\limits_{\Omega}}
\epsilon\nabla\varpi\nabla\widetilde{g}dx-%
{\displaystyle\int\limits_{\Omega}}
\epsilon\Delta\widetilde{g}\varpi dx+C(f,\widetilde{g})\\
& =J(\varpi;f)+%
{\displaystyle\int\limits_{\partial \Omega}}
\epsilon\frac{\partial\widetilde{g}}{\partial n}\varpi dx+C(f,\widetilde
{g})\\
& =J(\varpi;f)+C(f,\widetilde{g}),
\end{align*}
where the term $C(f,\widetilde{g})$ is considered as a constant during the
minimization process. Therefore, we have%
\begin{equation}
\arg\min_{v\in H_{0}^{1}(\Omega)}J(v,\widetilde{f})=\arg\min_{\varpi\in V_{g}%
}J(\varpi,f),
\end{equation}
where%
\begin{equation}
J(\varpi)=J(\varpi,f)=\frac{1}{2}%
{\displaystyle\int\limits_{\Omega}}
\left(  |\epsilon(x)\nabla\varpi|^{2}+\kappa(x)\varpi^{2}\right)  dx-%
{\displaystyle\int\limits_{\Omega}}
f\varpi dx.
\end{equation}

In practice, a penalty term can be added in the functional to enforce the
boundary condition, namely%
\begin{equation}
w=\arg\min_{\varpi\in H^{1}(\Omega)}J(\varpi)+\beta||\varpi-g||^{2}.
\end{equation}

In our numerical tests, the Ritz loss function is taken as
\begin{equation}
L_{\rm ritz}(u_{\theta})=\frac{1}{n}\sum_{\vx\in S}(\epsilon(\vx) |\nabla u_{\theta}(\vx)|^2/2+\kappa(\vx)u_{\theta}(\vx)^{2}/2-f(\vx)u_{\theta}(\vx) ) +\beta\frac{1}{\tilde{n}}\sum_{\vx\in \tilde{S}}(u_{\theta}(\vx)-{g}(\vx))^2,
\label{ritzlossnum}
\end{equation}
where $u_{\theta}(\vx)$ is the DNN output, $S$ is the sample set from $\Omega$ and $n$ is the sample size, $\tilde{n}$  indicates the number of sample set $\tilde{S}$ from $\partial\Omega$. We choose $\beta=1000$ for all experiments.

To see the learning accuracy, we also compute the $L^2$ error between $u_{\theta}(\vx)$ and $u_{\rm true}(\vx)$ on test data points $S_t=\{\vx_i\}_{i=1}^{n_t}$ inside the domain,
\begin{equation}
{\rm error}(u_{\theta}(\vx),u_{\rm true}(\vx))=\left (\frac{1}{n_{t}}\sum_{\vx\in S_t}(u_{\theta}(\vx)-u_{\rm true}(\vx))^2 \right )^{1/2}. \label{abserror}
\end{equation}

\section{Effectiveness of various MscaleDNN settings}
In this section, we will show that MscaleDNNs outperform normal fully-connected DNNs (indicated by ``normal" in the numerical results) in various settings, namely, the loss function of MscaleDNN decays faster to smaller values than that of normal fully-connected DNNs.It will also reflect smaller errors for the solutions for the MscaleDNN. First, we will carry out two test problems. Second, we will demonstrate that compact supported activation functions of  $\sReLU(x)$ and $\phi(x)$ are much better than the commonly used $\ReLU(x)$. Third, we use activation functions $\phi(x)$ to show MscaleDNN structures are better than normal fully connected one. Finally, we examine the effects of various scale selections.

\subsection{Two test problems}
To understand the performance of different MscaleDNNs and their parameters, here we consider one- and two- dimensional problems in fitting functions and solving PDEs,
and problems in 3-D in complex domains will be considered in the next section.

\paragraph{Test problem 1: Fitting problem}

\begin{figure}[htbp]
\centering
\subfloat[$d=1$]{\includegraphics[height=0.25\textheight]{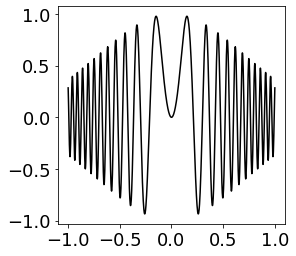}}
\subfloat[$d=2$]{\includegraphics[height=0.25\textheight]{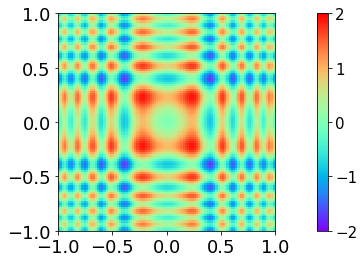}}
\caption{Test Problem 1: target functions for fitting problems.}
\label{func1}
\end{figure}

The target function for the fitting problem is $F: [-1, 1]^d \rightarrow \mathbb{R}$
\begin{equation}
F(\vx) = \sum_{j=1}^{d} g(x_{j})  \quad x_{j} \in [-1, 1],
\end{equation}
where $\vx=(x_1,\cdots,x_d)$,
$$ g(x) = e^{-x^2} \sin(\mu x^2). $$
In the case of $d=1$, we choose $\mu=70$ while for the case of $d=2$, $\mu=30$. The functions of $d=1$ and $d=2$ are shown in Fig. \ref{func1}. $5000$ training data at each epoch and $500$ test data are randomly sampled from $[-1, 1]^d$. All DNNs are trained by the Adam optimizer with learning rate $0.001$.

\paragraph{Test problem 2: Solving PB equations}
We will solve the elliptic equation (\ref{prob}) with $\epsilon=1$ and a constant $\kappa(\vx)=\lambda^2$ in a domain $\Omega = [-1, 1]^d$ and the right hand side
$$ f(\vx) = \sum_{i=1}^{d} (\lambda^2 + \mu^2) \sin(\mu x_i), $$
which gives a PB equation with the following exact solution,
$$ u(\vx) = \sum_{i=1}^{d} - \frac{\sin \mu}{\sinh \lambda} \sinh(\lambda x_i) + \sin(\mu x_i) $$ with corresponding  boundary condition given by the exact solution.

\begin{figure}[htbp]
\centering
\subfloat[$d=1$]{\includegraphics[height=0.25\textheight]{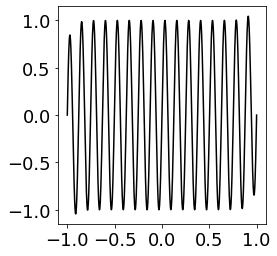}}
\subfloat[$d=2$]{\includegraphics[height=0.25\textheight]{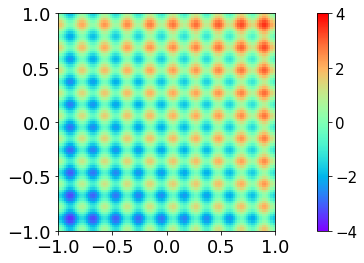}}
\caption{Test problem 2: exact solutions of 1-D and 2-D PB equation.}
\label{func2}
\end{figure}

For $d=1$, we choose $\lambda=20$, $\mu=50$. For $d=2$, we choose $\lambda=2$, $\mu=30$. The exact solutions for $d=1$ and $d=2$ are shown in Fig. \ref{func2}. DNNs are trained by Adam optimizer with learning rate $0.001$. $5000$ training data at each epoch are randomly sampled from $\Omega$. We choose the penalty coefficient for boundary as $\beta=1000$. The number of boundary data randomly sampled from $\partial\Omega$ is $400$ for $d=1$ and  $4000$ for $d=2$.

\subsection{Different activation functions}\label{exp1}
We use the following three network structures to examine the effectiveness of different activation functions by solving one-dimensional fitting and PDE problems described above:
\begin{enumerate}
\item fully-connected DNN with size \textbf{1-900-900-900-1} (\textbf{normal}).
\item MscaleDNN-1 with size \textbf{1-900-900-900-1} and scale coefficients of $\{1,2,4,8,16,32\}$ (\textbf{MscaleDNN-1(32)}).
\item MscaleDNN-2 with six subnetworks with size \textbf{1-150-150-150-1} and scale coefficients of $\{1,2,4,8,16,32\}$ (\textbf{MscaleDNN-2(32)}).
\end{enumerate}

\begin{figure}[htbp]
\centering
\subfloat[normal]{\includegraphics[width=0.3\linewidth]{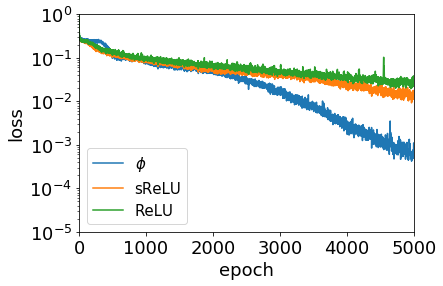}}
\subfloat[MscaleDNN-1]{\includegraphics[width=0.3\linewidth]{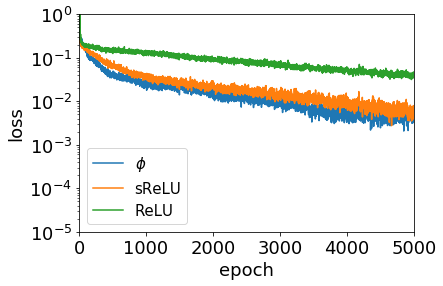}}
\subfloat[MscaleDNN-2]{\includegraphics[width=0.3\linewidth]{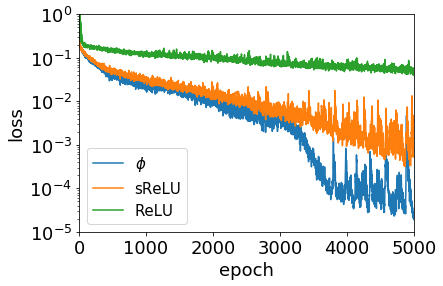}}
\caption{Different activation functions in 1-D fitting problems.}
\label{e1-1}
\end{figure}

\begin{figure}[htbp]
\centering
\subfloat[normal]{\includegraphics[width=0.3\linewidth]{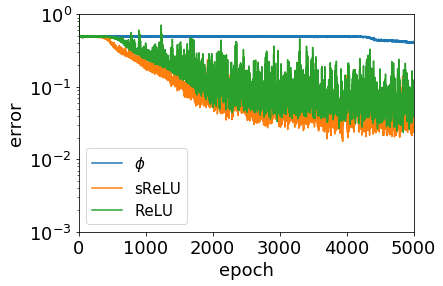}}
\subfloat[MscaleDNN-1]{\includegraphics[width=0.3\linewidth]{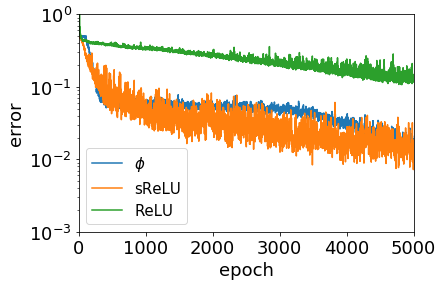}}
\subfloat[MscaleDNN-2]{\includegraphics[width=0.3\linewidth]{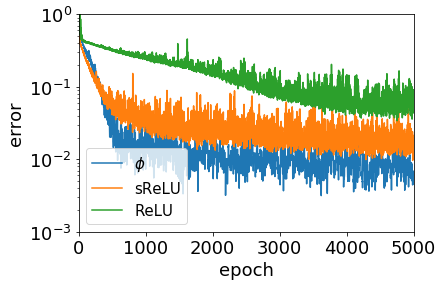}}
\caption{Different activation functions in a 1-D PB equation.}
\label{e1-2}
\end{figure}

\begin{figure}[htbp]
\centering
\subfloat[normal]{\includegraphics[width=0.3\linewidth]{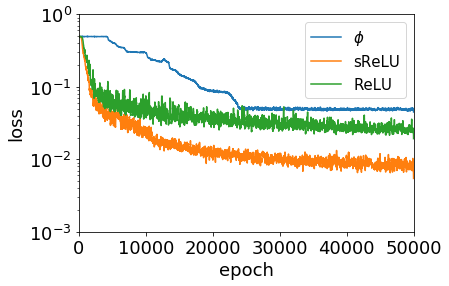}}
\subfloat[MscaleDNN-1]{\includegraphics[width=0.3\linewidth]{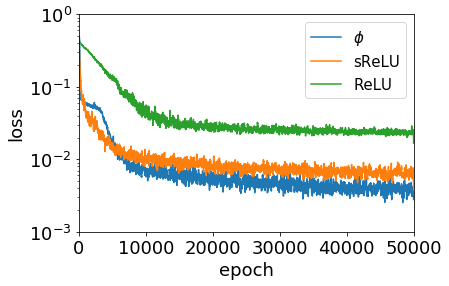}}
\subfloat[MscaleDNN-2]{\includegraphics[width=0.3\linewidth]{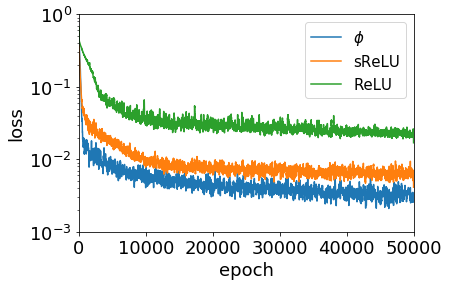}}
\caption{Different activation functions in a 1-D PB equation.}
\label{e1-3}
\end{figure}

In Fig. \ref{e1-3}, we increase the number of total epoch to 50000. The results are similar. Therefore, several thousand epochs are enough to compare the performance of networks.

We use three different activation functions, i.e., $\ReLU$, $\sReLU$, $\phi$ for the above structures. For normal network structure in the fitting problem, as shown in Fig. \ref{e1-1}(a), $\phi$ (blue) performs much better than other two activation functions. However, with a normal network structure to solve the PDE,  as shown in Fig. \ref{e1-2}(a), $\phi$ (blue) performs much worse than other two activation functions. The results indicate all three activation function are not stable in a normal fully connected structure. On the other hand, as shown in Fig. \ref{e1-1} (b, c),  and \ref{e1-2} (b, c), for both MscaleDNN structures, the performance of compact supported activation functions, $\sReLU$ (orange) and $\phi$ (blue), are much better than that of $\ReLU$ (green) for both test problems.

\subsection{Different network structures}\label{exp2}
In this subsection, we examine the effectiveness of the following different network structures with the activation function of $\phi(x)$:
\begin{enumerate}
\item fully-connected DNN with size \textbf{1-900-900-900-1} (\textbf{normal}).
\item MscaleDNN-1 with size \textbf{1-900-900-900-1} and scale coefficients of $\{1,2,4,8,16,32\}$ (\textbf{MscaleDNN-1(32)}).
\item MscaleDNN-2 with six subnetworks with size \textbf{1-150-150-150-1} and scale coefficients of $\{1,2,4,8,16,32\}$ (\textbf{MscaleDNN-2(32)}).
\end{enumerate}

\begin{figure}[htbp]
\centering
\subfloat[1d function]{\includegraphics[width=0.45\linewidth]{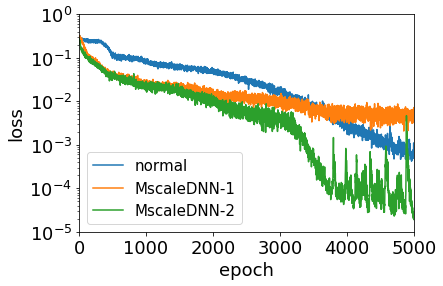}}
\subfloat[2d function]{\includegraphics[width=0.45\linewidth]{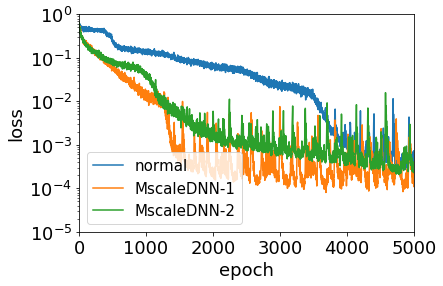}}
\caption{Different network structures in fitting problems.}
\label{e2-1}
\end{figure}

\begin{figure}[htbp]
\centering
\subfloat[1d equation]{\includegraphics[width=0.45\linewidth]{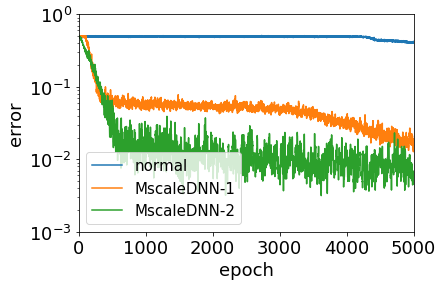}}
\subfloat[2d equation]{\includegraphics[width=0.45\linewidth]{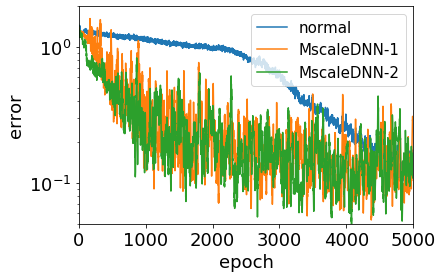}}
\caption{Different network structures in PDE problems.}
\label{e2-2}
\end{figure}

As shown in Fig. \ref{e2-1} and \ref{e2-2}, both MscaleDNN structures are better than normal structures in both problems. Two different MscaleDNN structures have similar performance in both test problems. As MscaleDNN-2 performs better than MscaleDNN-1 and also has much less connections compared with MscaleDNN-1 and a dynamic adaptive strategy of adding and removing scales can also be implemented, in the following we will use MscaleDNN-2 for further numerical experiments.

\subsection{Different scale selections in MscaleDNNs}\label{exp3}
In this subsection, we will test different scales for the activation function in MscaleDNNs:
\begin{enumerate}
\item fully-connected DNN with size \textbf{1-900-900-900-1} (\textbf{normal}).
\item MscaleDNN-2 with six subnetworks with size \textbf{1-150-150-150-1} and scale coefficients of $\{1,1,1,1,1,1\}$ (\textbf{MscaleDNN-2(1)}).
\item MscaleDNN-2 with three subnetworks with size \textbf{1-300-300-300-1} and scale coefficients of $\{1,2,3\}$ (\textbf{MscaleDNN-2(3)}).
\item MscaleDNN-2 with three subnetworks with size \textbf{1-300-300-300-1} and scale coefficients of $\{1,2,4\}$ (\textbf{MscaleDNN-2(4)}).
\item MscaleDNN-2 with six subnetworks with size \textbf{1-150-150-150-1} and scale coefficients of $\{1,2,3,4,5,6\}$ (\textbf{MscaleDNN-2(6)}).
\item MscaleDNN-2 with six subnetworks with size \textbf{1-150-150-150-1} and scale coefficients of $\{1,2,4,8,16,32\}$ (\textbf{MscaleDNN-2(32)}).
\end{enumerate}

\begin{figure}[htbp]
\centering
\subfloat[1d equation]{\includegraphics[height=0.2\textheight]{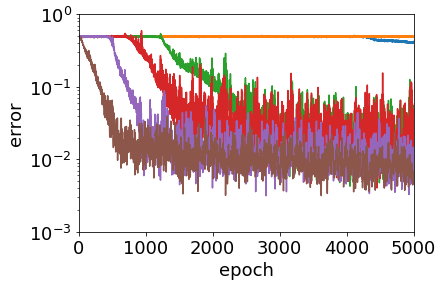}}
\subfloat[2d equation]{\includegraphics[height=0.2\textheight]{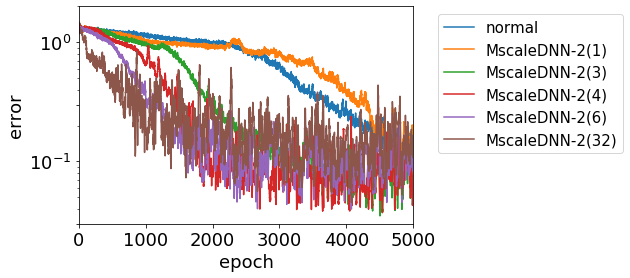}}
\caption{Different scale options in 1-D and 2-D PB equations.}
\label{e3-2}
\end{figure}

As shown in Fig. \ref{e3-2}, MscaleDNNs almost perform consistently better than normal DNNs. Note that with larger-range scale, MscaleDNN solves the problem faster. With all scales as $1$, the performance of DNN structure (MscaleDNN-2(1)) is much worse than those with multiscales in solving elliptic PDEs. Therefore, with the subnetwork structures with different scales, the MscaleDNN is able to achieve a faster convergence. These experiments show that MscaleDNNs with proper scales are more efficient in solving PDE problems and the selection of the scales are not too sensitive.

With these numerical experiments, we have demonstrated that MscaleDNN is much more efficient to solve elliptic PDEs and the preferred network is MscaleDNN-2 with the compact support function $\phi(x)$, which will be used for the rest of the paper for solving Poisson and PB equations in complex and/or singular domains.

\section{MscaleDNNs for Poisson and Poisson-Boltzmann equations in complex and singular domains}
In this section, we apply MscaleDNNs with activation function $\phi(x)$ to solve complex elliptic equations, including cases with a broad range of frequencies, variable coefficients, a ring-shaped domain, and a cubic domain with multiple holes. Finally, we apply the MscaleDNN to solve PB equations with geometric singularities, such as cusps and self-intersecting surfaces, which comes from a typical bead-model of bio-molecule. Through these experiments, we convincingly demonstrate that MscaleDNNs are an efficient and easy-implemented mesh-less method to solve complex elliptic PDEs.

\subsection{Poisson equation in complex domains}
\subsubsection{Broad range of frequencies}\label{exp4}

Consider the Poisson equation in $\Omega = [-1, 1]^d$,
\begin{equation}
-\Delta u(\vx) = f(\vx), \label{posvari}
\end{equation}
where
\begin{equation}
  f(\vx) = \sum_{i=1}^{d}  4 \mu^2 x_i^2 \sin(\mu x_i^2) - 2 \mu \cos(\mu x_i^2).
\end{equation}
The equation has an exact solution as
\begin{equation}
  u(\vx) = \sum_{i=1}^{d} \sin(\mu x_i^2),
\end{equation}
which will also provide the boundary condition in problem (\ref{posvari}).

In each training epoch, we sample $5000$ points inside the domain and $4000$ points from the boundary. We examine the following two structures:
\begin{enumerate}
\item  a fully-connected DNN with size \textbf{1-1000-1000-1000-1} (\textbf{normal}).
\item  a MscaleDNN-2 with five subnetworks with size \textbf{1-200-200-200-1}, and scale coefficients $\{1,2,4,8,16\}$. (\textbf{Mscale}).
\end{enumerate}

\begin{figure}[htbp]
\centering
\subfloat[exact]{\includegraphics[width=0.3\linewidth]{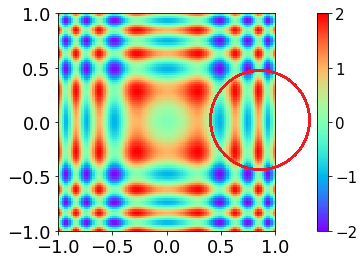}}
\subfloat[normal]{\includegraphics[width=0.3\linewidth]{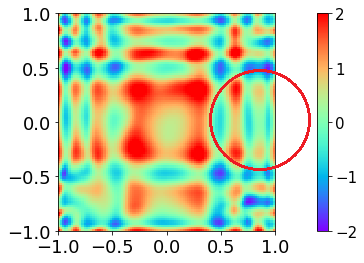}}
\subfloat[Mscale]{\includegraphics[width=0.3\linewidth]{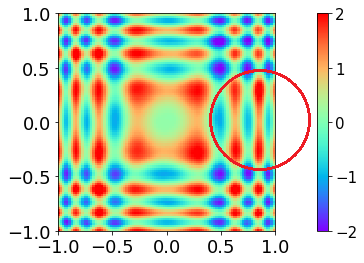}}
\caption{Two-dimensional case for problem (\ref{posvari}). As an example, the MscaleDNN well captures the oscillation in the red marked circle while the normal fully connected network fails.}
\label{func3}
\end{figure}

This problem does not have a fixed frequency but a broad range of frequencies. A commonly-used fully connected DNN will not be able to solve this problem. For $\mu=15$, the exact solution for the two-dimensional case of problem (\ref{posvari}) is shown in Fig. \ref{func3} (a) as a highly oscillated function. The solution, obtained by the normal DNN in Fig. \ref{func3} (b), fails to capture the oscillate structure, while the solution obtained by the MscaleDNN in Fig. \ref{func3} (c) captures well the different-scale oscillations. For example of the area marked by the red circle, the expected oscillation almost disappears in the solution of the normal networks while MscaleDNN solutions resolve the oscillations well. Similar behavior differences occur for the oscillations at four corners.

\begin{figure}[htbp]
\centering
\subfloat[2d Poisson equation]{\includegraphics[width=0.45\linewidth]{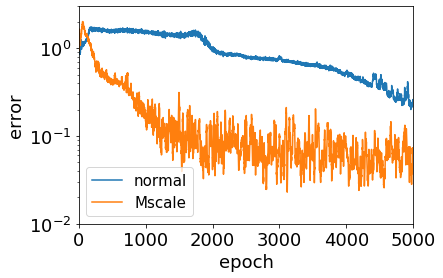}}
\subfloat[3d Poisson equation]{\includegraphics[width=0.45\linewidth]{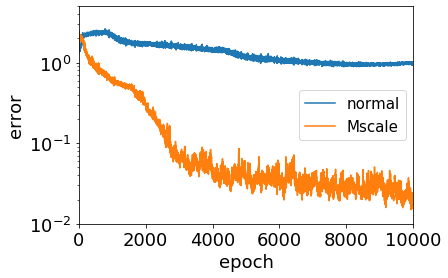}}
\caption{Error vs. epoch for problems with broad range of frequencies.}
\label{e4}
\end{figure}

The errors of the two-dimensional  and the three-dimensional problems are shown in Fig. \ref{e4} (a) and (b), respectively. In both cases, MscaleDNNs solve problems much faster to lower errors.

\subsubsection{A ring-shaped domain}\label{exp6}

Consider the Poisson equation (\ref{posvari}) in a ring-shaped domain $\Omega$ with its center at $(0,0)$ and  inner radius $1$ and outer radius $3$
with a source term
\begin{equation}
  f(\vx) = \mu^2 J_0(\mu |\vx - \vx_0|),
\end{equation}
where $J_0$ is the Bessel function.
The exact solution is given by
\begin{equation}
 u(\vx) = J_0(\mu |\vx - \vx_0|).
\end{equation}
Again, the boundary condition is given by the exact solution $u(\vx)$. We choose $\vx_0 = (0.5, 0)$ and solve the equation with $\mu=5$ and $\mu=10$.

\begin{figure}[htbp]
\centering
\subfloat[exact]{\includegraphics[width=0.3\linewidth]{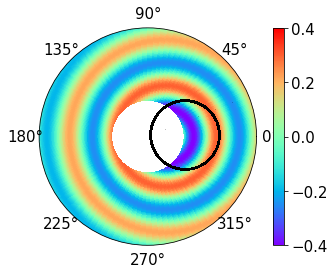}}
\subfloat[normal]{\includegraphics[width=0.3\linewidth]{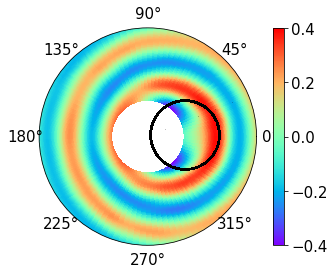}}
\subfloat[Mscale]{\includegraphics[width=0.3\linewidth]{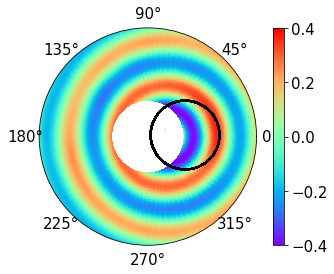}}
\caption{Exact and numerical solutions for the equation in a ring-shaped domain with $\mu=5$. The black circle is for illustration purpose only. }
\label{func4-m5}
\end{figure}

\begin{figure}[htbp]
\centering
\subfloat[exact]{\includegraphics[width=0.3\linewidth]{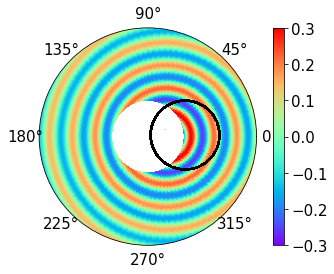}}
\subfloat[normal]{\includegraphics[width=0.3\linewidth]{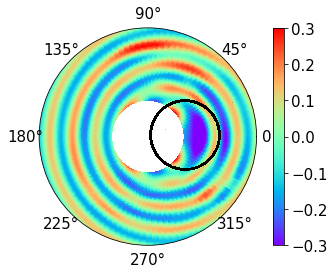}}
\subfloat[Mscale]{\includegraphics[width=0.3\linewidth]{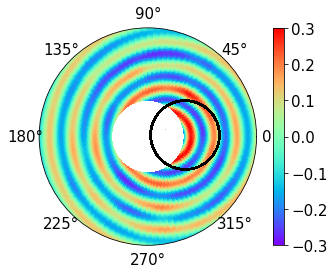}}
\caption{Exact and numerical solution for the equation in ring-shaped domain with $\mu=10$. The black circle is for illustration purpose only.}
\label{func4-m10}
\end{figure}

In each training epoch, we sample $5000$ points inside the domain and $4000$ points from the boundary. We examine the following two structures:
\begin{enumerate}
\item  a fully-connected DNN with size \textbf{1-500-500-500-1} (\textbf{normal}).
\item  a MscaleDNN-2 with five subnetworks with size \textbf{1-100-100-100-1} and scale coefficients $\{1,2,4,8,16\}$. (\textbf{Mscale}).
\end{enumerate}

\begin{figure}[htbp]
\centering
\subfloat[$\mu=5$]{\includegraphics[width=0.45\linewidth]{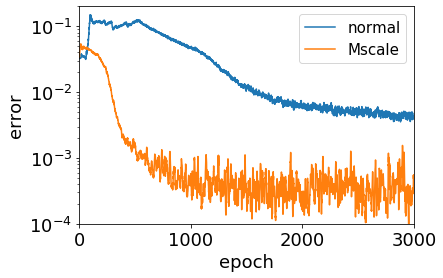}}
\subfloat[$\mu=10$]{\includegraphics[width=0.45\linewidth]{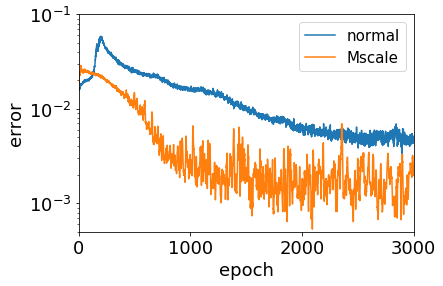}}
\caption{Error vs. epoch for the Poisson equation in ring-shaped domain.}
\label{e6}
\end{figure}

The exact solutions and numerical solutions obtained by normal and MscaleDNNs are shown in Fig. \ref{func4-m5} ($\mu=5$) and Fig. \ref{func4-m10}  ($\mu=10$). To highlight the superior performance of the MscaleDNNs, areas in the figures marked by the black circle show the region of the solution with the largest amplitude, the normal networks completely fail to capture the oscillations while the MscaleDNNs faithfully captures them in both cases. Again, as shown in Fig. \ref{e6},  MscaleDNN solves both problems with a much better accuracy.

\subsubsection{A square domain with a few holes}\label{exp7}

\noindent{\bf Domain one} The centers for three circle holes are $(-0.5, -0.5)$, $(0.5, 0.5)$, and $(0.5, -0.5)$, with radii of  $0.1$, $0.2$, and $0.2$, respectively. In each epoch, we randomly sample $3000$ on outer boundary, $800$ points on the boundary of each big hole and $400$ points on the boundary of the small hole.
\medskip

\noindent{\bf Domain two} The centers for three circle holes are $(-0.6, -0.6)$, $(0.3, -0.3)$ and $(0.6,0.6)$, with radii of $0.3$, $0.6$, $0.3$, respectively.  The boundary of the elliptic hole is described by $16(x + 0.5)^2 + 64(y - 0.5)^2 = 1$. The sample sizes at each epoch are $2400$, $1100$, $550$, and $400$ for the outer boundary, the boundary of the big circle hole, the boundary of each small circle hole, and the boundary of the elliptic hole, respectively.

We solve the Poisson equation (\ref{posvari}) with the source term as
\begin{equation}
f(\vx) = 2 \mu^2 \sin \mu x_1 \; \sin \mu x_2, \mu=7\pi.
\end{equation}
The exact solution is
\begin{equation}
u(\vx) = \sin \mu x_1 \sin \mu x_2.
\end{equation}
which also provides the boundary condition. In each training epoch, we sample $5000$ points inside the domain with the following two DNN structures:
\begin{enumerate}
\item  a fully-connected DNN with size \textbf{1-1000-1000-1000-1} (\textbf{normal}).
\item  a MscaleDNN-2 with five subnetworks with size \textbf{1-200-200-200-1}, and scale coefficients of $\{1,2,4,8,16\}$. (\textbf{Mscale}).
\end{enumerate}

\begin{figure}[htbp]
\centering
\subfloat[exact]{\includegraphics[width=0.3\linewidth]{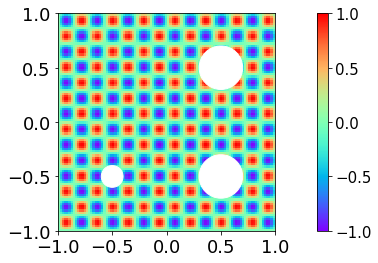}}
\subfloat[normal]{\includegraphics[width=0.3\linewidth]{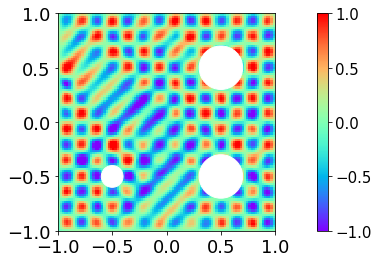}}
\subfloat[Mscale]{\includegraphics[width=0.3\linewidth]{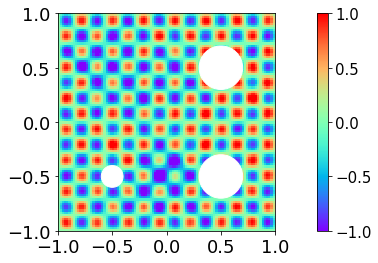}}
\caption{Exact and numerical solution for the Poisson equation in domain 1. }
\label{func5-r1}
\end{figure}

\begin{figure}[htbp]
\centering
\subfloat[exact]{\includegraphics[width=0.3\linewidth]{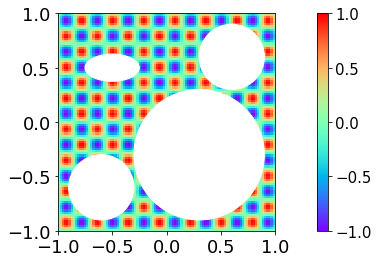}}
\subfloat[normal]{\includegraphics[width=0.3\linewidth]{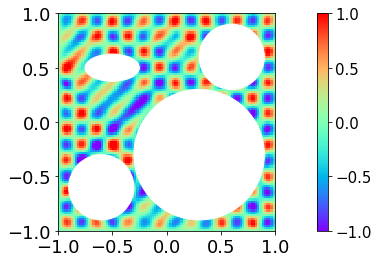}}
\subfloat[Mscale]{\includegraphics[width=0.3\linewidth]{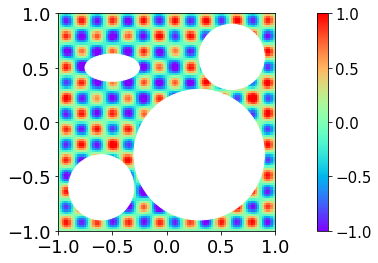}}
\caption{Exact and numerical solution for the Poisson equation in domain 2.}
\label{func5-r2}
\end{figure}

\begin{figure}[htbp]
\centering
\subfloat[domain 1]{\includegraphics[width=0.45\linewidth]{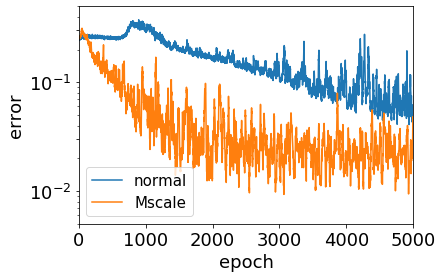}}
\subfloat[domain 2]{\includegraphics[width=0.45\linewidth]{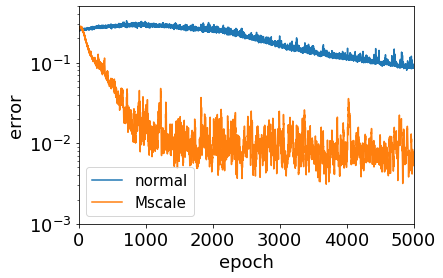}}
\caption{Error vs. epoch for the Poisson equation in square domains with few holes.}
\label{e7}
\end{figure}

As shown in Fig. \ref{e7}. MscaleDNNs solve both problems much faster to lower errors.

Compared with the exact solutions in Fig. \ref{func5-r1} (a) and Fig. \ref{func5-r2} (a), normal DNN fails to resolve the magnitidues of many oscillations as shown in Fig. \ref{func5-r1} (b) and Fig. \ref{func5-r2} (b) while MscaleDNNs capture each oscillation of the true solutions accurately as shown in Fig. \ref{func5-r1} (c) and Fig. \ref{func5-r2} (c).

\subsubsection{A square domain with many holes}\label{exp8}
\begin{figure}[htbp]
\centering
\includegraphics[width=0.45\linewidth]{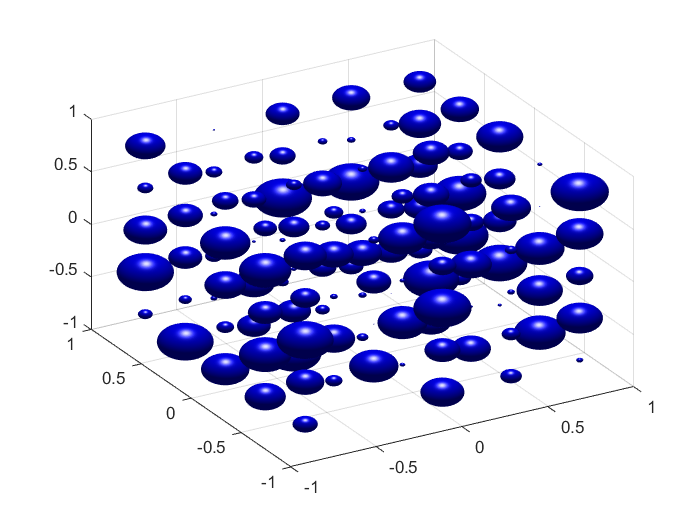}
\caption{Holes of domain for the problem}
\label{func6}
\end{figure}

To verify the capability of the MscaleDNN for complex domains, we consider a three dimensional cube $[-1, 1]^3$ with 125 holes inside removed as shown in Fig. \ref{func6}, and the holes are centered at a uniform mesh, i.e.,  $\{-0.8, -0.4, 0, 0.4, 0.8\}^3$, with radii randomly sampled from a uniform distribution in $[0,0.15]$. The sample sizes for training DNNs at each training epoch are $2500$ for the outer boundary and $1500$ for the inner holes ($12$ points for each hole).

Again, consider the Poisson equation with
$f(x)$ and the Dirichlet boundary condition given by the exact solution $ u(\vx)$ for the following three cases:
\begin{enumerate}
\item  Example 1: $ u(\vx) = \sin \mu x_1 \sin \mu x_2 \sin \mu x_3 $.
\item  Example 2: $ u(\vx) = {\mathrm{e}}^{\sin \mu x_1 + \sin \mu x_2 + \sin \mu x_3} $.
\item  Example 3: $ u(\vx) = \mathrm{e}^{ \sin \mu x_1 \sin \mu x_2 \sin \mu x_3 } $.
\end{enumerate}
\begin{figure}[htbp]
\centering
\subfloat[Example 1]{\includegraphics[width=0.3\linewidth]{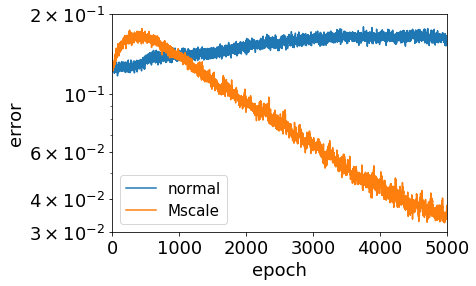}}
\subfloat[Example 2]{\includegraphics[width=0.3\linewidth]{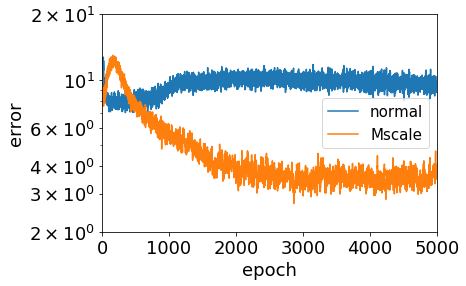}}
\subfloat[Example 3]{\includegraphics[width=0.3\linewidth]{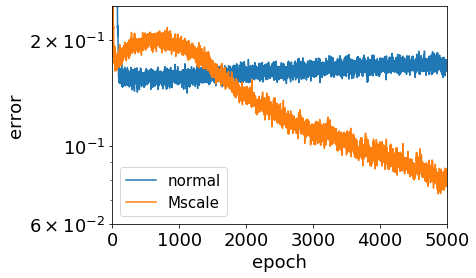}}
\caption{Error vs. epoch for the PDEs in square domain with many holes.}
\label{e8}
\end{figure}
The difficulty of this problem consists of the complex holes and oscillatory exact solutions with $\mu=7\pi$. In each training epoch, we sample $5000$ points inside the domain, and compare the following two structures:
\begin{enumerate}
\item  a fully-connected DNN with size \textbf{1-1000-1000-1000-1} (\textbf{normal}).
\item  a MscaleDNN-2 with five subnetworks with size \textbf{1-200-200-200-1}, and scale coefficients of $\{1,2,4,8,16\}$. (\textbf{Mscale}).
\end{enumerate}

As shown in Fig. \ref{e8} for all three cases, the normal fully-connected structures do not converge for such complex problems at all while MscaleDNNs can solve the problem with much smaller errors.

\subsection{Poisson-Boltzmann equations with domain and source singularities}

\subsubsection {Variable coefficients}\label{exp5}

Consider the PB equation (\ref{prob}) in $\Omega = [-1, 1]^3$ with
\begin{equation}
 f(\vx) = (\mu_1^2 + \mu_2^2 + \mu_3^2 + x_1^2 + 2 x_2^2 + 3 x_3^2) \sin(\mu_1 x_1) \sin(\mu_2 x_2) \sin(\mu_3 x_3),
\end{equation}
and
\begin{equation}
    \kappa(\vx) = (x_1^2 + 2 x_2^2 + 3 x_3^2),
\end{equation}
  which has an exact solution as
  \begin{equation}
u(\vx) = \sin(\mu_1 x_1) \sin(\mu_2 x_2) \sin(\mu_3 x_3).
\end{equation}
The boundary condition is given by the exact solution $u(\vx)$. We choose $\mu_1=15, \mu_2=20, \mu_3=25$.
\begin{figure}[htbp]
\centering
\includegraphics[width=0.55\linewidth]{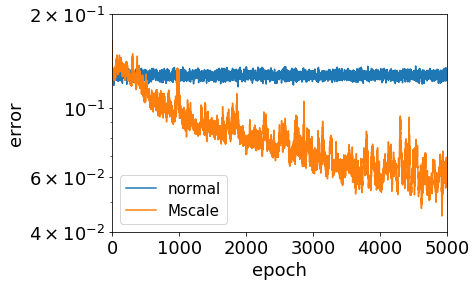}
\caption{Error vs. epoch for variable coefficient PB equation.}
\label{e5}
\end{figure}
In each training epoch, we sample $5000$ points inside the domain and $4000$ points from the boundary. We compare the following two DNN structures:
\begin{enumerate}
\item  a fully-connected DNN with size \textbf{1-900-900-900-1} (\textbf{normal}).
\item  a MscaleDNN-2 with six subnetworks with size \textbf{1-150-150-150-1} and scale coefficients $\{1,2,4,8,16,32\}$. (\textbf{Mscale}).
\end{enumerate}

As shown in Fig. \ref{e5}, during the training process, the error of the MscaleDNN decays significantly, while the error of the normal DNN almost keeps unchanged. Therefore, MscaleDNN solves the problem much faster with a much better accuracy.

\subsubsection{Geometric singularities}\label{exp9}
In this subsection, we consider the PB equation (\ref{prob}) in a domain with geometric singularities and jump condition on interior interfaces, which arises from the simulation of solvation of bio-molecules. Consider an open bounded domain $\Omega_1 \subset \mathbb{R}^3$, which divides $\mathbb{R}^3$ into two disjoint open subdomains by the surface $\Gamma = \partial \Omega_1$. $\Omega_1$ is identified as the bio-molecule, and $\Omega_2=\mathbb{R}^3 \setminus \Omega_1$ is the solvent region. The exact solution $u(x)$ is also divided into two parts, $u_1(x)$ is defined in $\Omega_1$ and $u_2(x)$ in $\Omega_2$. The solution will also satisfy the transmission condition (\ref{continuouscond}) (\ref{derivativecond}) along the interface $\Gamma$ and a decaying condition at the $\infty$, i.e.
\begin{equation}
\lim_{|\vx| \rightarrow\infty} u_2(\vx) = 0. \label{boundarycond}
\end{equation}

\begin{figure}[htbp]
\centering
\includegraphics[width=0.40\linewidth]{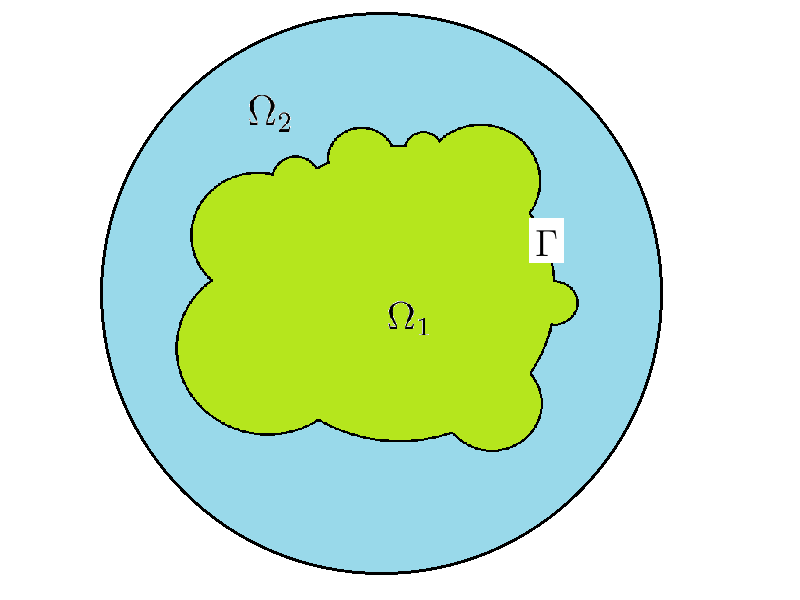}
\includegraphics[width=0.45\linewidth]{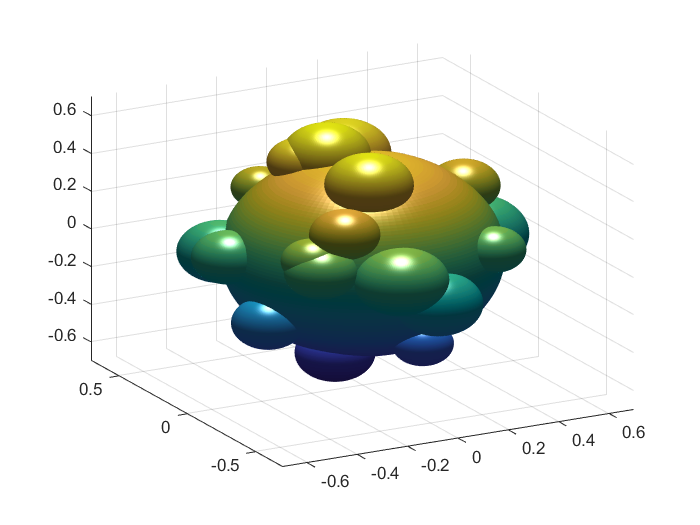}
\caption{Solution domain: (left) truncation of computation domain, (right) geometric singularity.}
\label{region}
\end{figure}

To deal with the unbounded domain, we truncate the solution domain to a large  ball or cube, denoted by $\Omega$ satisfying $\Omega_1 \subset \Omega$ and we re-define $\Omega_2=\Omega \setminus \Omega_1$ and set an approximate condition $u_2=0$ on the boundary of the ball (Fig. \ref{region} (left) ) and such a crude boundary condition will surely introduce error to the PDEs solution. Higher order boundary conditions have been studied extensively, and as we are more interested in the performance of the DNNs near the interior interface, we will not ponder over this issue here.

The domain with geometric singularities is constructed as follows. We choose a big ball with a center at $(0,0,0)$ and a radius of $0.5$.  $20$ points are randomly selected  on the surface of the big ball as the centers of small balls. Radiuses of the small balls are randomly sampled from $[0.1, 0.2]$. $\Omega_1$ is the union of these balls and the big ball. The shape of $\Omega_1$ is illustrated in Fig. \ref{region} (right). The intersections among balls cause geometric singularities, such as kinks, which poses major challenges for obtaining mesh generation for traditional finite element and boundary element methods and accurate solution procedures.
\%label{reg3d}

Following two examples are considered. In both examples, coefficients $\epsilon(\vx)$ and $\kappa(\vx)$ are chosen as piece-wise constant. Singular sources for the PB equations, which can occur from the point charges inside  bio-molecules or ions in the solvents, will be considered later. These point charge sources, modeled by Dirac delta function, will create point singularity in the solution, which can be removed by subtracting a singular solution \cite{chern2003}.

\paragraph{Example 1} The exact solution is
\begin{equation}
 u(\vx) = \frac{e^{\sin \mu x_1 + \sin \mu x_2 + \sin \mu x_3}}{|\vx|^2 + 1} (|\vx|^2 - 1)
 \end{equation}
with coefficients for the PB equation as
\begin{equation} \mu = 15, \ \epsilon(\vx) = 1, \ \kappa(\vx) = 1 \ {\rm for}\ \vx\in\Omega_1, \ \epsilon(\vx) = 1, \kappa(\vx) = 5 \ {\rm for}\ \vx\in\Omega_2.
\end{equation}
The whole domain is truncated by a ball with center at $(0,0,0)$ and a radius $1$ with zero boundary condition on the sphere.

\paragraph{Example 2} We choose
\begin{equation}
f(\vx) = \frac{e^{\sin \mu x_1 + \sin \mu x_2 + \sin \mu x_3}}{|\vx|^2 + 1} (|\vx|^2 - 1)
\end{equation}
with coefficients
\begin{equation}
    \mu = 20, \ \epsilon(\vx) = 1 \ {\rm for}\ \vx\in\Omega_1, \  \epsilon(\vx) = 80 \ {\rm for}\ \vx\in\Omega_2,\ \kappa(\vx) = 1.
\end{equation}

In this case, the computational domain is obtained with a truncation by a cube $[-1,1]^3$ and the reference solution is calculated by finite difference method (FDM) with a sufficient fine mesh ensuring enough accuracy.

In example 1, in each training epoch, we sample $5000$ points inside the domain $\Omega$ and $4000$ points on boundary $\partial \Omega$. In example 2, we sample $6000$ points inside the domain $\Omega$, $3000$ points on boundary $\partial \Omega$. We train MscaleDNNs with the Ritz loss function in (\ref{ritzlossnum}). Note that the continuity condition in (\ref{continuouscond}) is satisfied since we use a single network to fit the whole domain $\Omega$; The natural condition in (\ref{derivativecond}) is also automatically satisfied due to the use of the Ritz loss.

\begin{figure}[htbp]
\centering
\subfloat[loss]{\includegraphics[width=0.45\linewidth]{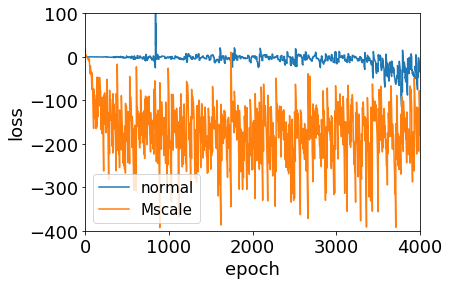}}
\subfloat[relative error]{\includegraphics[width=0.45\linewidth]{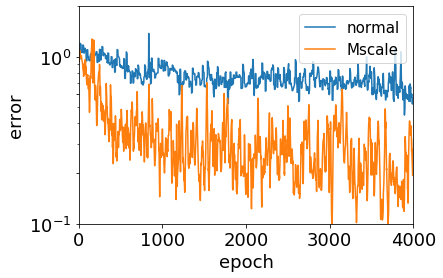}}
\caption{Loss and relative error vs. epoch for the PB equation in a domain with geometric singularities. (Example 1)}
\label{e9}
\end{figure}
\begin{figure}[htbp]
\centering
\subfloat[loss]{\includegraphics[width=0.45\linewidth]{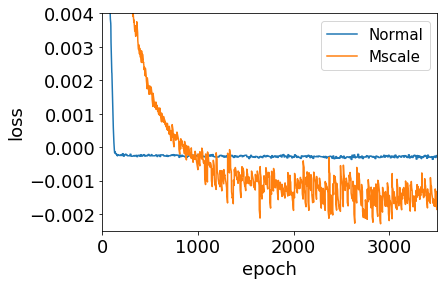}}
\subfloat[relative error]{\includegraphics[width=0.43\linewidth]{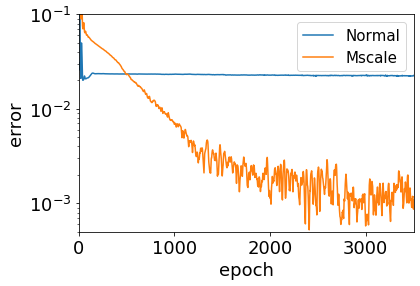}}
\caption{Loss and relative error vs. epoch for the PB equations in a domain with geometric singularities. (Example 2)}
\label{e14}
\end{figure}

We examine the following two structures:
\begin{enumerate}
\item a fully-connected DNN with size \textbf{1-1000-1000-1000-1} (\textbf{normal}).
\item  a MscaleDNN-2 with five subnetworks with size \textbf{1-200-200-200-1}, and scale coefficients of $\{1,2,4,8,16\}$. (\textbf{Mscale}).
\end{enumerate}
\begin{figure}[htbp]
\centering
\includegraphics[width=0.55\linewidth]{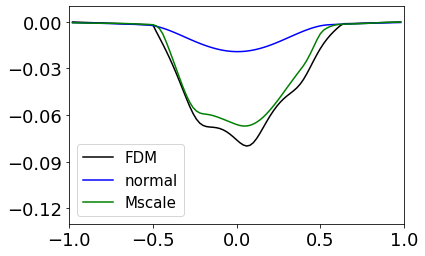}
\caption{Numerical solutions of Example 2 on line $x_1 = x_3 = 0$.}
\label{e9f}
\end{figure}

 Since the value of the exact solution is small, we show the relative $L^2$ error for both cases.
 As in practice, the exact solution is unknown,  therefore, we also show the training loss for both examples, which could be used as a possible criteria to terminate the training. For example 1 as shown in Fig. \ref{e9}, the training loss in Fig. \ref{e9}(a) and the error in Fig. \ref{e9}(b) have similar trends, that is, the MscaleDNN converge faster to smaller values, compared with the normal DNN. For example 2 shown in Fig. \ref{e14}, the MscaleDNN shows a similar advantage over the normal DNN. These examples indicate that with by just monitoring the training loss,  MscaleDNN solves the PB equations with non-smooth solution over singular domains much faster and with better accuracy.

For illustration, we show a cross section of the solution in the second example. The reference solution is obtained by the FDM. Numerical solutions on the line $x_1 = x_3 = 0$ obtained by FDM($h=0.02$), normal DNN($5000$ epochs) and MscaleDNN($5000$ epochs) are shown in Fig. \ref{e9f}. The output of the normal fully connected network gives a wrong solution in the interior of the singular domain while the MscaleDNN gives a satisfactory approximation to the reference solution.

\subsubsection{Source and geometric singularities}\label{expn}
\begin{figure}[htbp]
\centering
\includegraphics[width=0.35\linewidth]{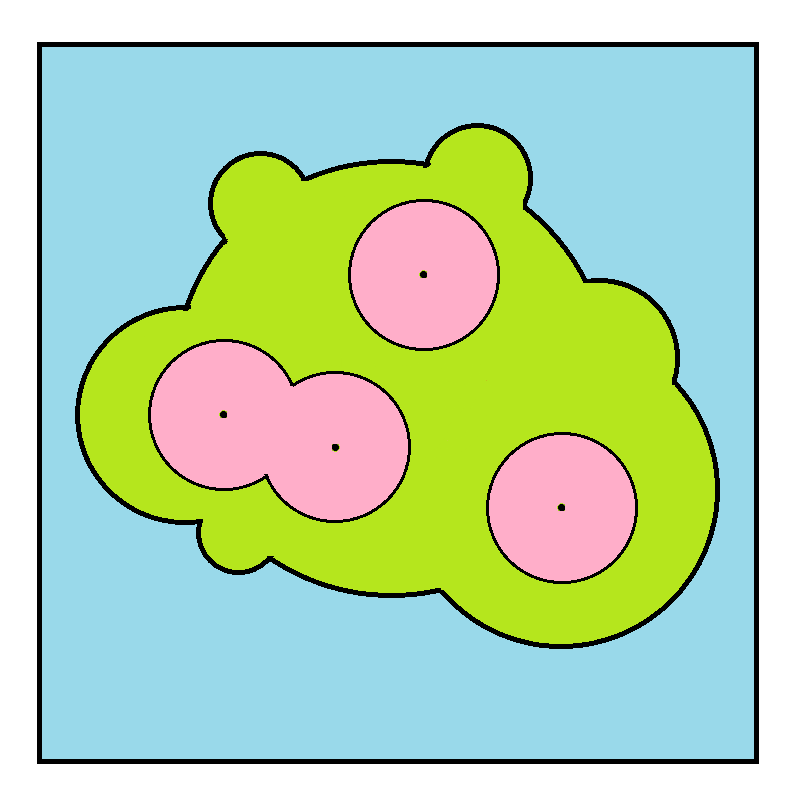}
\caption{Spherical truncation of the physical domain.}
\label{region2}
\end{figure}

In this subsection, we consider the PB equation (\ref{prob}) with singular sources, that is,
\begin{equation}
f(\vx) = \sum_{k=1}^{K} q_k \delta(\vx - \mathbf{s}_k),
\label{singsource}
\end{equation}
where $\delta(x)$ is Dirac delta function, $q_k$ and $\mathbf{s}_k$ represent the charge  and position of one nuclei in the bio-molecule, respectively. We assume that the distance between nucleus and the molecule interface is bigger than a constant $R_0$, that is, $\Omega_0 = \{\vx : \exists k, |\vx - \mathbf{s}_k| < R_0 \} \subset \Omega_1$. In Fig. \ref{region2}, the blue part represents the solvent domain $\Omega_2$, the green part represents the biomolecular domain $\Omega_1 \setminus \Omega_0$, and the pink part represents $\Omega_0$, which contains all charges.

To deal with singularities, we define
\begin{equation}
\bar{u}(\vx) = \sum_{k=1}^{K} q_k G(\vx - \mathbf{s}_k) m(\vx - \mathbf{s}_k),
\end{equation}
where
\begin{equation}
G(\vx) = \frac{1}{4 \pi \epsilon_1} \frac{e^{-\frac{\kappa_1}{\sqrt{\epsilon_1}} |\vx|}}{|\vx|}
\end{equation}
and the mollifier function
\begin{align}
\begin{cases}
m(\vx)= 1 - \left(\frac{|\vx|}{R_0}\right)^3 \left(4 - 3 \frac{|\vx|}{R_0} \right), &\quad |\vx| < R_0, \\
m(\vx)= 0. &\quad |\vx| > R_0.
\end{cases}
\end{align}
By above definitions, it can be verified easily that $\bar{u}(\vx)$ satisfies
\begin{align}
\begin{cases}
-\Delta \bar{u}(\vx) + \kappa^2 \bar{u}(\vx) = \sum_{k=1}^{K} q_k \delta(\vx - \mathbf{s}_k) + \sum_{k=1}^{K} q_k F(|\vx - \mathbf{s}_k|), &\qquad \vx \in \Omega_0, \\
\bar u= \frac{\partial \bar u}{\partial n}=0, &\qquad \vx \in \partial \Omega_0, \\
\bar{u}(\vx) = 0, &\qquad \vx \in \Omega_0^c.
\end{cases}
\end{align}
Next, we define
\begin{equation}
    w(x) = u(x) - \bar{u}(x)  \chi_{\Omega_0}(x),
\end{equation}
which will satisfy the following equations without singularities
\begin{equation}
\label{eqws}
  - \epsilon(x) \triangle w(x) + \kappa^2(x) w(x)  = f(x) \chi_{\Omega_0}(x),
\end{equation}
where
\begin{equation}
f(x) = -\sum_{k=1}^{K} q_k F(|x - \mathbf{s}_k|),
\end{equation}
\begin{align}
\begin{cases}
F(r) = \frac{3 \mathrm{e}^{-\frac{\kappa_1}{\sqrt{\epsilon_1}} r}}{\pi R_0^4} (2 R_0 - 3 r + 2 \frac{\kappa_1}{\sqrt{\epsilon_1}} r^2 - 2 R_0 \frac{\kappa_1}{\sqrt{\epsilon_1}} r), &\quad r < R_0\\
F(r) = 0,& \quad r > R_0.
\end{cases}
\end{align}

We will present the numerical results for equation (\ref{eqws}).

\paragraph{Example 1}
In the first example, we choose $\Omega = [-1,1]^3$. $\Omega_1$ is a ball with center $(0,0,0)$ and radius $R = 0.7$. Parameters are chosen as
$$ \mathbf{s} = (0,0,0), \quad q = 1, \quad R_0 = 0.5, \quad \epsilon(\vx) = 1 \ {\rm for}\ \vx\in\Omega_1, \ \epsilon(\vx) = 80 \ {\rm for}\ \vx\in\Omega_2,\ \kappa(\vx) = 0. $$

The exact solution is
\begin{equation}
    u(x) = \frac{1}{4\pi |x| \epsilon_1} - (\frac{1}{\epsilon_1} - \frac{1}{\epsilon_2}) \frac{1}{4\pi R}, \quad \vx\in\Omega_1; \qquad u(x) = \frac{1}{4\pi |x| \epsilon_2, }, \quad \vx\in\Omega_2,
\end{equation}
and, correspondingly
\begin{align*}
\begin{cases}
w(x) = \frac{1}{4 \pi \epsilon_1 |x|} \left(\frac{|x|}{R_0}\right)^3 \left(4 - 3 \frac{|x|}{R_0} \right) - (\frac{1}{\epsilon_1} - \frac{1}{\epsilon_2}) \frac{1}{4\pi R}, &\quad |x| < R_0, \\
w(x) = \frac{1}{4\pi |x| \epsilon_1} - (\frac{1}{\epsilon_1} - \frac{1}{\epsilon_2}) \frac{1}{4\pi R}, &\quad R_0 < |x| < R, \\
w(x) = \frac{1}{4\pi |x| \epsilon_2}, &\quad |x| > R.
\end{cases}
\end{align*}
\paragraph{Example 2}
In the second example, we choose $\Omega = [-1,1]^3$. The domain is constructed as follows. We choose a large ball with center $(0,0,0)$ and radius $0.7$. $20$ points are randomly selected on the surface of the large ball as the centers of small balls. Radii of the small balls are randomly sampled from $[0.1, 0.3]$. $\Omega_1$ is the union of these balls.

The singular source term in (\ref{singsource}) is constructed as follows. The position of each charge is randomly selected in the ball with center $(0,0,0)$ and radius $0.5$ and the quantity of charges is from $[-0.5, 0.5]$. We choose $R_0 = 0.2$. Parameters are chosen as
$$ \epsilon(\vx) = 1 \ {\rm for}\ \vx\in\Omega_1, \ \epsilon(\vx) = 80 \ {\rm for}\ \vx\in\Omega_2,\ \kappa(\vx) = 0. $$
The reference solution is again calculated by a FDM with a very fine mesh.
\paragraph{DNN results} In each training epoch, we sample $6000$ points inside the domain $\Omega$ and $3000$ points on boundary $\partial \Omega$. In example 1, we examine the following two structures:
\begin{enumerate}
\item  fully-connected DNN with size \textbf{1-1000-1000-1000-1} (\textbf{normal}).
\item  MscaleDNN-2 with five subnetworks with size \textbf{1-200-200-200-1}, and scale coefficients of $\{1,2,4,8,16\}$ (\textbf{Mscale}).
\end{enumerate}

In example 2, the equation is more complex than before, we need more neurons to approximate the complex solution. In example 2, we examine the following two structures with boundary penalty $\beta=100$:
\begin{enumerate}
\item  fully-connected DNN with size \textbf{1-1500-1000-1000-500-1} (\textbf{normal}).
\item  MscaleDNN-2 with five subnetworks with size \textbf{1-300-200-200-100-1}, and scale coefficients of $\{1,2,4,8,16\}$ (\textbf{Mscale}).
\end{enumerate}

\begin{figure}[htbp]
\centering
\subfloat[Example 1]{\includegraphics[width=0.45\linewidth]{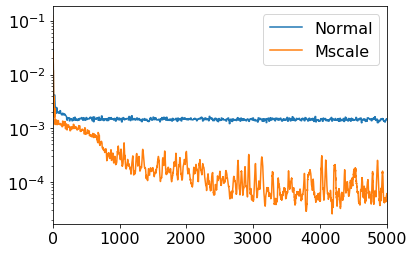}}
\subfloat[Example 2]{\includegraphics[width=0.45\linewidth]{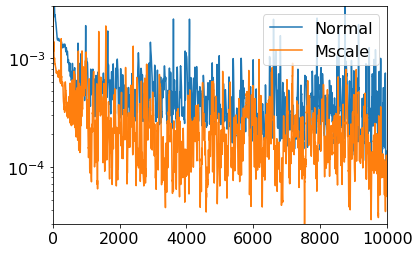}}
\caption{Error vs. epoch for the PDEs in domain with geometric and source singularities.}
\label{enl}
\end{figure}

As shown in Fig. \ref{enl}, the errors of the MscaleDNN decays much faster and achieves  much smaller errors after training for both examples.

\begin{figure}[htbp]
\centering
\subfloat[Example 1]{\includegraphics[width=0.45\linewidth]{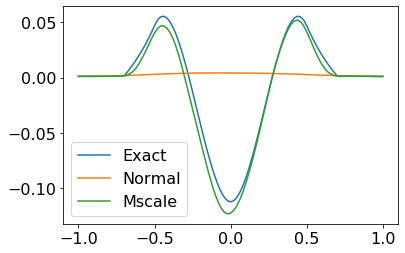}}
\subfloat[Example 2]{\includegraphics[width=0.45\linewidth]{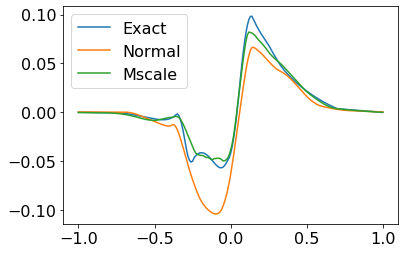}}
\caption{Numerical solutions on line $x_2 = x_3 = 0$.}
\label{ens}
\end{figure}

The numerical solutions on the line $x_2 = x_3 = 0$ obtained by the FDM ($h=0.01$), normal DNN ($10000$ epochs) and MscaleDNN ($10000$ epochs) are shown in Fig. \ref{ens}. The output of the normal fully connected network can not capture the peaks in exact solution very well.

\begin{figure}[htbp]
\centering
\subfloat[FDM]{\includegraphics[width=0.33\linewidth]{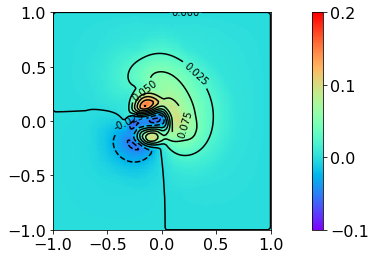}}
\subfloat[Normal]{\includegraphics[width=0.33\linewidth]{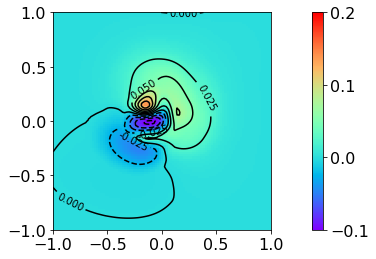}}
\subfloat[Mscale]{\includegraphics[width=0.33\linewidth]{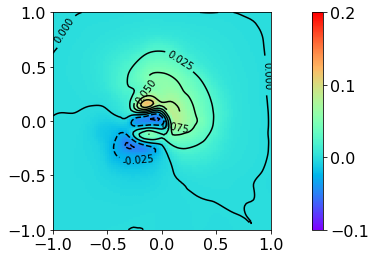}}
\caption{Numerical solutions of example 2 on plane $x_3 = 0$.}
\label{ens1}
\end{figure}

\begin{figure}[htbp]
\centering
\subfloat[Normal]{\includegraphics[width=0.45\linewidth]{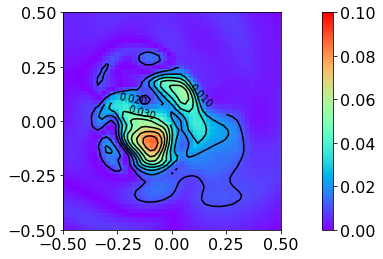}}
\subfloat[Mscale]{\includegraphics[width=0.45\linewidth]{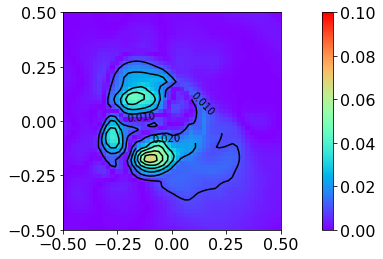}}
\caption{Errors of example 2 on plane $x_3 = 0$.}
\label{ens2}
\end{figure}

For the second example, the numerical solutions and errors on the surface $x_3 = 0$ around the bio-molecule obtained by FDM ($h=0.01$), normal DNN ($10000$ epochs) and MscaleDNN ($10000$ epochs) are shown in Fig. \ref{ens1} and Fig. \ref{ens2}.
\section{Conclusion and future work}
In this paper, we have introduced a new kind of multi-scale DNNs, using a frequency domain scaling technique and compactly supported activation functions,  to generate a multi-scale capability for finding the solutions of elliptic PDEs with rich frequency contents. By using a radial scaling in the Fourier domain of the solutions, the MscaleDNN is shown to be an efficient mesh-less and easy-to-implement method for PDEs on complex and singular domains, for which solvers by finite element and finite difference methods may be costly due to the need of mesh generations and solution of large linear systems.

For future work, we will also explore the idea of activation function with the mother wavelet properties as proposed in \cite{MsDNNarxiv}, which should give further frequency localization and separation capability in the MscaleDNNs. Applications of the MscaleDNN to large scale computational engineering problems will be carried out, especially,  in comparison with finite element and finite difference methods. More importantly,  an area to be explored is to apply the MscaleDNN to  high dimensional PDEs such as Schrodinger equations for many body quantum systems, issues of high dimensional sampling and low dimensional structure of solutions will be studied.

\medskip
\noindent{\bf Acknowledgments}

W.C. is supported by US National Science Foundation (Grant No. DMS-1950471). Z.X. is supported by National Key R\&D Program of China (2019YFA0709503), and Shanghai Sailing Program.

\bibliographystyle{plain}
\bibliography{DLref}

\begin{thebibliography}{10}

\bibitem{holst2001}
NA~Baker, Joseph~S Sept~D, Holst MJ, and McCammon JA.
\newblock Electrostatics of nanosystems: application to microtubules and the
  ribosome.
\newblock {\em Proceedings of the National Academy of Sciences},
  98(18):10037--41, 2001.

\bibitem{basri2019convergence}
Ronen Basri, David Jacobs, Yoni Kasten, and Shira Kritchman.
\newblock The convergence rate of neural networks for learned functions of
  different frequencies.
\newblock {\em arXiv preprint arXiv:1906.00425}, 2019.

\bibitem{cai2013}
Wei Cai.
\newblock {\em Computational Methods for Electromagnetic Phenomena,
  electrostatics in solvation, scatterings, and electron transport}.
\newblock Cambirdge University Press, 2013.

\bibitem{cai2019phasednn}
Wei Cai, Xiaoguang Li, and Lizuo Liu.
\newblock A phase shift deep neural network for high frequency approximation
  and wave problems.
\newblock {\em to appear in SIAM J. Scientific Computing, arXiv:1909.11759},
  2019.

\bibitem{MsDNNarxiv}
Wei Cai and Zhi-Qin~John Xu.
\newblock Multi-scale deep neural networks for solving high dimensional pdes.
\newblock {\em Arxiv preprint, arXiv:1910.11710}, 2019.

\bibitem{cao_towards_2020}
Yuan Cao, Zhiying Fang, Yue Wu, Ding-Xuan Zhou, and Quanquan Gu.
\newblock Towards {Understanding} the {Spectral} {Bias} of {Deep} {Learning}.
\newblock {\em arXiv:1912.01198 [cs, stat]}, 2020.

\bibitem{chern2003}
I.-L. Chern, J.-G. Liu, and W.-C. Wang.
\newblock Accurate evaluation of electrostatics for macromolecules in solution.
\newblock {\em Meth. Appl. Anal.}, 10:309–328, 2003.

\bibitem{daubechies1992ten}
Ingrid Daubechies.
\newblock {\em Ten lectures on wavelets}, volume~61.
\newblock Siam, 1992.

\bibitem{deng2018learning}
Mo~Deng, Shuai Li, and George Barbastathis.
\newblock Learning to synthesize: splitting and recombining low and high
  spatial frequencies for image recovery.
\newblock {\em arXiv preprint arXiv:1811.07945}, 2018.

\bibitem{weinan2017deep}
Weinan E, Jiequn Han, and Arnulf Jentzen.
\newblock Deep learning-based numerical methods for high-dimensional parabolic
  partial differential equations and backward stochastic differential
  equations.
\newblock {\em Communications in Mathematics and Statistics}, 5(4):349--380,
  2017.

\bibitem{weinan2018deep}
Weinan E and Bing Yu.
\newblock The deep ritz method: A deep learning-based numerical algorithm for
  solving variational problems.
\newblock {\em Communications in Mathematics and Statistics}, 6(1):1--12, 2018.

\bibitem{hamilton2019dnn}
A~Hamilton, T~Tran, MB~Mckay, B~Quiring, and PS~Vassilevski.
\newblock Dnn approximation of nonlinear finite element equations.
\newblock Technical report, Lawrence Livermore National Lab.(LLNL), Livermore,
  CA (United States), 2019.

\bibitem{han2018solving}
Jiequn Han, Arnulf Jentzen, and E~Weinan.
\newblock Solving high-dimensional partial differential equations using deep
  learning.
\newblock {\em Proceedings of the National Academy of Sciences},
  115(34):8505--8510, 2018.

\bibitem{han2018deep}
Jiequn Han, Linfeng Zhang, Roberto Car, et~al.
\newblock Deep potential: A general representation of a many-body potential
  energy surface.
\newblock {\em Communications in Computational Physics}, 23(3), 2018.

\bibitem{he2018relu}
Juncai He, Lin Li, Jinchao Xu, and Chunyue Zheng.
\newblock Relu deep neural networks and linear finite elements.
\newblock {\em arXiv preprint arXiv:1807.03973}, 2018.

\bibitem{kingma2014adam}
Diederik~P Kingma and Jimmy Ba.
\newblock Adam: A method for stochastic optimization.
\newblock {\em arXiv preprint arXiv:1412.6980}, 2014.

\bibitem{liao2019deep}
Yulei Liao and Pingbing Ming.
\newblock Deep nitsche method: Deep ritz method with essential boundary
  conditions.
\newblock {\em arXiv preprint arXiv:1912.01309}, 2019.

\bibitem{Lindskog}
S.~Lindskog.
\newblock Structure and mechanism of carbonic anhydrase.
\newblock {\em Pharmacol. Therapeut.}, 74:1--20, 1997.

\bibitem{luo2019theory}
Tao Luo, Zheng Ma, Zhi-Qin~John Xu, and Yaoyu Zhang.
\newblock Theory of the frequency principle for general deep neural networks.
\newblock {\em arXiv preprint arXiv:1906.09235}, 2019.

\bibitem{pan2018learning}
Jinshan Pan, Sifei Liu, Deqing Sun, Jiawei Zhang, Yang Liu, Jimmy Ren, Zechao
  Li, Jinhui Tang, Huchuan Lu, Yu-Wing Tai, et~al.
\newblock Learning dual convolutional neural networks for low-level vision.
\newblock In {\em Proceedings of the IEEE conference on computer vision and
  pattern recognition}, pages 3070--3079, 2018.

\bibitem{rahaman2018spectral}
Nasim Rahaman, Aristide Baratin, Devansh Arpit, Felix Draxler, Min Lin, Fred
  Hamprecht, Yoshua Bengio, and Aaron Courville.
\newblock On the {Spectral} {Bias} of {Neural} {Networks}.
\newblock In {\em International {Conference} on {Machine} {Learning}}, pages
  5301--5310, 2019.

\bibitem{raissi2019physics}
Maziar Raissi, Paris Perdikaris, and George~E Karniadakis.
\newblock Physics-informed neural networks: A deep learning framework for
  solving forward and inverse problems involving nonlinear partial differential
  equations.
\newblock {\em Journal of Computational Physics}, 378:686--707, 2019.

\bibitem{shaham2018provable}
Uri Shaham, Alexander Cloninger, and Ronald~R Coifman.
\newblock Provable approximation properties for deep neural networks.
\newblock {\em Applied and Computational Harmonic Analysis}, 44(3):537--557,
  2018.

\bibitem{strofer2019data}
Carlos~Michelen Strofer, Jin-Long Wu, Heng Xiao, and Eric Paterson.
\newblock Data-driven, physics-based feature extraction from fluid flow fields
  using convolutional neural networks.
\newblock {\em Communications in Computational Physics}, 25(3):625--650, 2019.

\bibitem{wang2020mesh}
Zhongjian Wang and Zhiwen Zhang.
\newblock A mesh-free method for interface problems using the deep learning
  approach.
\newblock {\em Journal of Computational Physics}, 400:108963, 2020.

\bibitem{wu2020multigrid}
Chao-Yuan Wu, Ross Girshick, Kaiming He, Christoph Feichtenhofer, and Philipp
  Krahenbuhl.
\newblock A multigrid method for efficiently training video models.
\newblock In {\em Proceedings of the IEEE/CVF Conference on Computer Vision and
  Pattern Recognition}, pages 153--162, 2020.

\bibitem{xu2019frequency}
Zhi-Qin~John Xu, Yaoyu Zhang, Tao Luo, Yanyang Xiao, and Zheng Ma.
\newblock Frequency principle: Fourier analysis sheds light on deep neural
  networks.
\newblock {\em Accepted by Communications in Computational Physics,
  arXiv:1901.06523}, 2019.

\bibitem{xu_training_2018}
Zhi-Qin~John Xu, Yaoyu Zhang, and Yanyang Xiao.
\newblock Training {Behavior} of {Deep} {Neural} {Network} in {Frequency}
  {Domain}.
\newblock In {\em Neural {Information} {Processing}}, Lecture {Notes} in
  {Computer} {Science}, pages 264--274, 2019.

\bibitem{wei2007}
W.~Yu, S.and~Geng and G.W. Wei.
\newblock Treatment of geometric singularities in implicit solvent models.
\newblock {\em J. Chem. Phys.}, 126:244108, 2007.

\bibitem{zhang2019explicitizing}
Yaoyu Zhang, Zhi-Qin~John Xu, Tao Luo, and Zheng Ma.
\newblock Explicitizing an implicit bias of the frequency principle in
  two-layer neural networks.
\newblock {\em arXiv preprint arXiv:1905.10264}, 2019.

\end{thebibliography}

\end{document}